\documentclass[onecolumn,aps,prd,preprintnumbers,showpacs,superscriptaddress,nofootinbib,amsmath,amssymb,floats,floatfix,showkeys,notitlepage,longbibliography]{revtex4-1}

\usepackage{comment}
\usepackage{orcidlink}
\usepackage{lipsum}
\usepackage{graphicx}
\usepackage{subfigure}
\usepackage{palatino}
\usepackage{sans}
\usepackage{hyperref}
\hypersetup{colorlinks=true,linkcolor=blue,urlcolor=blue,citecolor=blue}
\usepackage[toc,page]{appendix}
\usepackage[normalem]{ulem}
\usepackage{adjustbox}
\usepackage{latexsym}
\usepackage{amsmath}
\usepackage{amssymb}
\usepackage{amsfonts}
\usepackage{dcolumn}
\usepackage{bm}
\usepackage{tikz}
\usepackage{bigints}
\usepackage{array,tabularx,multirow,booktabs}
\usepackage[tracking=true]{microtype}
\usepackage{soul} 
\SetTracking{}{500}
\SetTracking{encoding={*}, shape=sc}{40}
\UseRawInputEncoding 
\allowdisplaybreaks

\newcommand{\be}{\begin{align}}
\newcommand{\ee}{\end{align}}

\def\be{\begin{equation}}
\def\ee{\end{equation}}
\def\bestar{\begin{equation*}}
\def\eestar{\end{equation*}}

\newcommand{\bea}{\begin{align}}\newcommand{\eea}{\end{align}}
\newcommand{\brr}{\begin{array}}\newcommand{\err}{\end{array}}
\newcommand{\bit}{\begin{itemize}}\newcommand{\eit}{\end{itemize}}
\newcommand{\ben}{\begin{enumerate}}\newcommand{\een}{\end{enumerate}}

\newcommand{\ba}{\begin{array}}
\newcommand{\ea}{\end{array}}

\graphicspath{{./images/} }

\newcolumntype{M}[1]{>{\centering\arraybackslash}m{#1}}
\newcolumntype{N}{@{}m{0pt}@{}}

\newcounter{sxn}

\newcounter{axn}

\newdimen\mybaselineskip
\newcommand{\beeq}{\begin{equation}}
\newcommand{\eneq}{\end{equation}}
\newcommand{\beqn}{\begin{align}}
\newcommand{\eeqn}{\end{align}}

\newcommand{\beal}{\setcounter{letter}{1} \begin{align}}
\newcommand{\eeal}{\addtocounter{equation}{1} \end{align}}

\newcommand{\larrow}{\,\,\,\,\hbox to 30pt{\rightarrowfill}
\,\,\,\,}
\newcommand{\slarrow}{\,\,\,\hbox to 20pt{\rightarrowfill}
\,\,\,}

\def\la{\raise.16ex\hbox{$\langle$}\lower.16ex\hbox{}  }
\def\ra{\, \raise.16ex\hbox{$\rangle$}\lower.16ex\hbox{} }

\def\psibar{ \psi \kern-.65em\raise.6em\hbox{$-$} \lower.6em\hbox{} }
\def\psibarb{ \psi \kern-.65em\raise.6em\hbox{$-$}  }

\begin{document} 

\title{Shadow and quasinormal modes of the rotating
Einstein-Euler-Heisenberg black holes}
 
\author{Gaetano Lambiase
\orcidlink{0000-0001-7574-2330}}
\email{lambiase@sa.infn.it}
\affiliation{Dipartimento di Fisica ``E.R Caianiello'', Università degli Studi di Salerno, Via Giovanni Paolo II, 132 - 84084 Fisciano (SA), Italy.}
\affiliation{Istituto Nazionale di Fisica Nucleare - Gruppo Collegato di Salerno - Sezione di Napoli, Via Giovanni Paolo II, 132 - 84084 Fisciano (SA), Italy.}

\author{Dhruba Jyoti Gogoi \orcidlink{0000-0002-4776-8506}}
\email{moloydhruba@yahoo.in}
\affiliation{Department of Physics, Moran College, Moranhat, Charaideo 785670, Assam, India.}
\affiliation{Theoretical Physics Division, Centre for Atmospheric Studies, Dibrugarh University, Dibrugarh
786004, Assam, India.}

\author{Reggie C. Pantig
\orcidlink{0000-0002-3101-8591}}
\email{rcpantig@mapua.edu.ph}
\affiliation{Physics Department, Map\'ua University, 658 Muralla St., Intramuros, Manila 1002, Philippines}

\author{Ali \"Ovg\"un
\orcidlink{0000-0002-9889-342X}
}
\email{ali.ovgun@emu.edu.tr}
\affiliation{Physics Department, Eastern Mediterranean University, Famagusta, 99628 North Cyprus via Mersin 10, Turkiye.}

\begin{abstract}
The Einstein-Euler-Heisenberg (EEH) black hole model is an extension of classical black hole solutions in general relativity, incorporating quantum electrodynamics (QED) effects via the Euler-Heisenberg Lagrangian. The Euler-Heisenberg Lagrangian describes the nonlinear corrections to Maxwell's equations due to virtual electron-positron pair production in a strong electromagnetic field. When this Lagrangian is coupled with Einstein's field equations, it leads to modified black hole solutions that take into account these quantum corrections. In this paper, we investigate the impact of the black hole charge $Q_e$ on the properties of the rotating and electrically charged Einstein-Euler-Heisenberg black holes (EEH). To this aim, we analyzed and discussed findings as to how the black hole charge $Q_e$ affects certain black hole properties such as null regions, shadow cast and its observables, and quasinormal modes (QNMs) relative to the Kerr and Kerr-Newman cases. We find that the presence of a screened charge due to the associated QED effects in this screened Maxwell theory might noticeably alter the properties of black holes, offering insights into the interplay between gravity and quantum field effects.
\end{abstract}

\date{\today}


\maketitle


\section{Introduction} \label{intro}

The Event Horizon Telescope (EHT) Collaboration's work on capturing the first images of supermassive black holes is a monumental achievement in astrophysics. The EHT is a global radio telescope network that works together to form a virtual Earth-sized telescope using the Very Long Baseline Interferometry (VLBI) technique. By synchronizing multiple telescopes across different continents, the EHT can achieve the high resolution necessary to observe details as small as the event horizon of a black hole. The image of M87*, released in April 2019, was the first direct visual evidence of a black hole's existence \cite{EventHorizonTelescope:2019dse,EventHorizonTelescope:2019ths, EventHorizonTelescope:2022xqj}. It confirmed theoretical predictions based on General Relativity and provided new insights into the behavior of matter and light around supermassive black holes, first analyzed through the seminal works of Synge and Luminet \cite{Synge:1966okc, Luminet:1979nyg} for static black holes, and Bardeen \cite{Bardeen:1973tla} for the axisymmetric case. Then, the image of Sgr. A*, released in May 2022, provided the first visual confirmation of the black hole at the center of our galaxy. It also validated the methods developed for M87* and underscored the dynamic nature of Sgr A* \cite{EventHorizonTelescope:2022wkp,EventHorizonTelescope:2022wok}. EHT's success has profound implications for astrophysics as it opens up new avenues for studying black holes' environments, testing gravity theories under extreme conditions, and understanding the fundamental physics of accretion and jet formation. The precise shape of these shadows encodes critical physical parameters, such as the black hole’s mass and spin, and the study of black hole shadows has proven instrumental in addressing fundamental questions across a broad spectrum of topics \cite{Vagnozzi:2022moj}, including the behavior of accretion disks \cite{Uniyal:2022vdu,Zeng:2020dco,Zeng:2021dlj}, the nature of dark matter/energy \cite{Pantig:2022whj,Pantig:2022sjb,Zeng:2020vsj},  modified gravity theories \cite{Ovgun:2020gjz,Kuang:2022xjp,Mustafa:2022xod,Kumaran:2022soh,Cimdiker:2021cpz,Okyay:2021nnh,Atamurotov:2022knb,Olmo:2023lil,Asukula:2023akj,Zeng:2023zlf}, and the existence of extra dimensions \cite{Vagnozzi:2019apd}. These intriguing questions have ignited a surge of theoretical and experimental research into black hole shadows.

The Einstein-Euler-Heisenberg system is considered an effective action of a superstring theory, where static and spherically symmetric black hole solutions were constructed in Ref. \cite{Yajima:2000kw}, and the electric charge was included \cite{Ruffini:2013hia}. The charged static case was then extended in Ref. \cite{Breton:2019arv} to an axisymmetric case using the Newman-Janis algorithm. Then, its rotating structure was studied \cite{Breton:2022fch, Zhong:2021mty}. Time-like particle motion was considered in Ref. \cite{Amaro:2023ull}, and many studies about the non-rotating case of EEH black holes exist in the literature \cite{Zeng:2022pvb,Breton:2021mju,Luo:2022gdz,Dai:2022mko,Feng:2022otu,Maceda:2018zim,Maceda:2020rpv,Rehman:2023hro,Mushtaq:2024cap, Hu:2020usx}. One of the aims of this paper is to explore the properties of the rotating EEH black hole through the shadow cast, shadow radius, and observables.

Black holes, intriguing celestial entities governed by Einstein's theory of gravity, represent enigmatic phenomena in the universe. A seminal moment in the study of black holes occurred with the detection of Gravitational Waves (GWs) on September 14th, 2015 \cite{PhysRevLett.116.061102}. This milestone not only deepened our understanding of black holes but also paved the way for experimental tests of gravitational theories. According to Einstein's theory of general relativity, GWs originate from the acceleration of massive objects, causing disturbances in the fabric of spacetime. These waves carry essential information about the dynamics and kinematics of the astronomical sources that produce them. Advanced instruments such as LIGO and Virgo have played pivotal roles in detecting GWs. When two black holes merge, they coalesce into a final black hole that emits GWs exhibiting distinct wave patterns known as ring-down modes. These GWs manifest quasinormal modes (QNMs) that depend on the mass and spin of the resulting black hole. Analyzing GW data using these QNMs is crucial for unveiling the mysterious properties of black holes and acquiring valuable insights into their nature.

QNMs represent a significant and captivating aspect of black hole physics. They signify the oscillations of a black hole that gradually attenuate over time, characterized by intricate frequencies. Termed "quasinormal" because they are not precisely normal modes, which would perpetually oscillate \cite{Vishveshwara:1970zz, Press:1971wr,Kokkotas:1999bd}, they fade away due to dissipative mechanisms such as gravitational wave emission. QNMs are complex values that portray the emission of gravitational waves from compact, massive celestial objects in the cosmos. The real component of QNMs indicates the emission frequency, while the imaginary component corresponds to their decay rate. Understanding QNMs is imperative as they encode vital information about the attributes of black holes, including their mass, angular momentum, and the characteristics of the surrounding spacetime \cite{Li:2021zct}. Moreover, delving into QNMs offers insights into the nature of black holes and the strong gravitational regime, which is challenging to explore using alternative methodologies. These modes play a fundamental role in comprehending the structure and evolution of black holes and their involvement in astrophysical phenomena such as gravitational wave signals. Recent years have witnessed extensive research into the exploration of GWs and the QNMs displayed by black holes within various modified gravity theories \cite{Rincon:2018ktz, Liu:2022ygf, Rincon:2021gwd, Anacleto:2021qoe, Lambiase:2023hng, sekhmani_electromagnetic_2023, Gogoi:2023kjt, Parbin:2022iwt, karmakar_quasinormal_2022,Gogoi:2022wyv,Gogoi:2021cbp,Gogoi:2021dkr,Pantig:2022gih, Gogoi:2024scc,Gogoi:2024vcx,Gogoi:2023lvw}.

 \textcolor{black}{Motivated by the need to bridge theoretical insights with observational tests, we extend the established framework of the EEH theory beyond its traditional presentation. While previous studies (e.g., Breton et al. \cite{Breton:2019arv}) provide a study of the theory and the derivation of rotating black hole solutions, our work focus on the observational test of  black hole charge $Q_e$ in EEH theory. This parameter not only enriches the theoretical foundation but also allows us to explore new phenomenological implications. In our study, we derive the theoretical shadow radius of the black hole and further impose constraints on the shadow angular radius. Furthermore, our study extends beyond the analysis of the static and rotating black hole geometries by also exploring their dynamical response through QNMs which describe the damped oscillations of perturbations in the black hole spacetime and serve as unique fingerprints of the black hole’s properties, such as mass, spin, and charge. These modes are not only theoretically significant—providing insights into the stability and structure of the black hole—but they are also observationally relevant, as they can be directly compared with gravitational wave signals detected by current observatories. These original contributions set our work apart by not only revisiting the established theory but also by paving the way for future experimental validations.}

The program of the paper is as follows: In Sect. \ref{sec2}, we give a brief review of the EEH theory and the EEH rotating black hole and we explore the null regions such as the event horizon and ergosphere. Sect. \ref{sec3} examines the null geodesic, and in Sect. \ref{sec4}, we study the different observables such as the shadow cast, shadow radius, etc. Sect. \ref{sec5} examines the QNMs, both the rotating and static cases. Finally, we form a conclusion and state possible future research directions. The paper use the metric signature $(-,+,+,+)$ and geometrized units by applying $G = c = 1$.

\section{Brief review of Einstein-Euler-Heisenberg theory and EEH rotating black hole} \label{sec2}

In this section, we briefly review the Einstein-Euler-Heisenberg theory by the action, expressed as follows \cite{Gibbons:2001sx,Breton:2019arv}:

\begin{equation}
    S = \frac{1}{16 \pi G} \int_{\mathcal{M}} d^4x \sqrt{-g} R + \frac{1}{4 \pi} \int_{\mathcal{M}} d^4x \sqrt{-g} \left( -X + \frac{A}{2} X^2 + \frac{B}{2} Y^2 \right),
\end{equation}
Here, \( R \) represents the Ricci scalar curvature, and \( G \) is Newton's constant, which we set to \( G = 1 \). The quantities \( X \) and \( Y \) are the two independent relativistic invariants and pseudo-invariants formed from the Maxwell field in four dimensions: 
$   X = -\frac{1}{4} F_{\mu \nu} F^{\mu \nu}, \quad Y = \frac{1}{4} F_{\mu \nu} \, {}^*F^{\mu \nu}.$
  \textcolor{black}{Here, $A = \frac{8 \alpha^2}{45 m^4}$ and $B = \frac{7 \alpha^2}{180 m^4} = \frac{7}{4} A$, where $m$ is the electron mass, and $\alpha$ is the fine structure constant.} ${}^*F_{\mu \nu}$ is the dual of the Faraday tensor $F_{\mu \nu} = A_{\mu, \nu} - A_{\nu, \mu}$, it is defined as
$
{}^*F_{\mu \nu} = \frac{1}{2} \epsilon_{\mu \nu \rho \sigma} F^{\rho \sigma}, \quad \epsilon_{\mu \nu \rho \sigma} \text{ is the completely antisymmetric tensor that satisfies } \epsilon_{\mu \nu \rho \sigma} \epsilon^{\mu \nu \rho \sigma} = -4!.$

The Lagrangian density for the Euler–Heisenberg non-linear electrodynamics is given by \( L(X, Y) = -X + \frac{A}{2} X^2 + \frac{B}{2} Y^2 \). It is important to note that \( P_{\mu \nu} \) coincides with \( F_{\mu \nu} \) in the case of linear Maxwell theory. In the general case, it is 
expressed as:

\begin{equation}
    P_{\mu \nu} = -(L_X F_{\mu \nu} + L_Y {}^*F_{\mu \nu}),
\end{equation}
where subscripts on $L$ denote differentiation 
,so it reduces to
\begin{equation}
    P_{\mu \nu} = -\left( (1 + A X) F_{\mu \nu} + B Y {}^*F_{\mu \nu} \right).
\end{equation}

The two independent invariant and pseudo-invariant quantities, \( s \) and \( t \), are denoted in terms of the dual Plebanski variables \( P_{\mu \nu} \), and are defined as follows:
\begin{equation}
    s = -\frac{1}{4} P_{\mu \nu} P^{\mu \nu}, \quad t = -\frac{1}{4} P_{\mu \nu} \, {}^*P^{\mu \nu},
\end{equation}
with ${}^*P_{\mu \nu} = \frac{1}{2} \epsilon_{\mu \nu \rho \sigma} P^{\rho \sigma}$.
We can write the covariant Hamiltonian \( H(s, t) \) as
\begin{equation}
    H(s, t) = -\frac{1}{2} P^{\mu \nu} F_{\mu \nu} - L.
\end{equation}


 \textcolor{black}{In the Euler--Heisenberg theory, one typically expands the Hamiltonian \(H(s,t)\) up to first order in the small (dimensionless) quantities \((A,s)\) and \((B,t)\). Although \(A\) and \(B\) themselves have dimensions of \(\text{[mass]}^{-4}\), the products \((A,s)\) and \((B,t)\) are dimensionless if \(s\) and \(t\) are chosen appropriately, for further details, see Ref.~\cite{Breton:2019arv}. Consequently, the Hamiltonian takes the form}
\begin{equation}
    H(s, t) = s - \frac{A}{2} s^2 - \frac{B}{2} t^2.
\end{equation}
 \textcolor{black}{where the terms proportional to \((A,s^2)\) and \((B,s^2)\) represent the leading QED corrections in the weak-field. }

We now write the equations of motion for the coupled system as \cite{Salazar:1987ap}
\begin{equation}
    D_\mu P^{\mu \nu} = 0, \quad R_{\mu \nu} - \frac{1}{2} R g_{\mu \nu} = 8 \pi T_{\mu \nu},
\end{equation}
and the energy-momentum tensor
\begin{equation}
    T_{\mu \nu} = \frac{1}{4 \pi} \left[ H_s P_{\mu}{}^\beta P_{\nu \beta} + g_{\mu \nu} \left( 2 s H_s + t H_t - H \right) \right].
\end{equation}
We define the energy-momentum tensor for the Euler–Heisenberg non-linear electromagnetic field as follows:
\begin{equation}
    T_{\mu \nu} = \frac{1}{4 \pi} \left[ (1 - A s) P_{\mu}{}^\beta P_{\nu \beta} + g_{\mu \nu} \left( s - \frac{3}{2} A s^2 - \frac{B}{2} t^2 \right) \right]
\end{equation}
\begin{equation}
    = \frac{1}{4 \pi} \left[ (-1 - 2 A X) F_{\mu}{}^\beta F_{\nu \beta} - B Y \left( F_{\mu} {}^*F^{\beta} + {}^*F_{\mu} F_{\nu \beta} \right) + g_{\mu \nu} \left( s - \frac{3}{2} A s^2 - \frac{B}{2} t^2 \right) \right].
\end{equation}
Note that $A = B = 0$ gives the standard Maxwell energy-momentum tensor.

Then, we use $F_{\mu \nu}$ with $P_{\mu \nu}$, with
\begin{equation}
    F_{\mu \nu} = \left[ H_s + H_t \right] P_{\mu \nu} = (1 - A s - B t) P_{\mu \nu},
\end{equation}
To find a solution, the following static and spherically symmetric black hole metric is considered:
\begin{equation}
    ds^2 = -\left( 1 - \frac{2m(r)}{r} \right) dt^2 + \left( 1 - \frac{2m(r)}{r} \right)^{-1} dr^2 + r^2 \left( d\theta^2 + \sin^2 \theta d\phi^2 \right).
\end{equation}

We assume the following ansatz for the Plebanski dual variables of the non-linear electromagnetic field in the electrically charged case:
\begin{equation}
P_{\mu \nu}=\frac{Q_{e}}{r^{2}}\left(\delta_{\mu}^{0} \delta_{\nu}^{1}-\delta_{\mu}^{1} \delta_{\nu}^{0}\right),
\end{equation}
 \textcolor{black}{where $Q_{e}$ is the black hole electric charge \cite{Breton:2019arv}.}

We obtain the result from the component \((0,0)\) of Einstein's equations
\begin{equation}
m(r)=M-\frac{Q_{e}^{2}}{2 r}+A \frac{Q_{e}^{4}}{40 r^{5}}
\end{equation}
Therefore, we write the electrically charged static black hole solution as \cite{Breton:2019arv}
\begin{eqnarray}
\mathrm{d} s^{2}=-\left(1-\frac{2 M}{r}+\frac{Q_{e}^{2}}{r^{2}}-A \frac{Q_{e}^{4}}{20 r^{6}}\right) \mathrm{d} t^{2}+\left(1-\frac{2 M}{r}+\frac{Q_{e}^{2}}{r^{2}}-A \frac{Q_{e}^{4}}{20 r^{6}}\right)^{-1} \mathrm{~d} r^{2}+r^{2}\left(\mathrm{~d} \theta^{2}+\sin ^{2} \theta \mathrm{d} \phi^{2}\right)    
\end{eqnarray}

Subsequently, it is derived Lorentz covariant solutions for spinning systems, which represent a relativistic version of the Newman-Janis technique. These solutions are obtained as follows \cite{Gurses:1975vu, Breton:2019arv}:

\begin{align}\mathrm{d} s^2 & =-\left(1-\frac{2 M r-Q_e^2+A Q_e^4 / 20 r^4}{\rho^2}\right) \mathrm{d} t^2+\frac{\rho^2}{\Delta} \mathrm{d} r^2-\frac{4 a\left(M r-\frac{Q_e^2}{2}+A Q_e^4 / 40 r^4\right) \sin ^2 \theta}{\rho^2} \mathrm{~d} t \mathrm{~d} \varphi \nonumber \\& +\rho^2 \mathrm{~d} \theta^2+\frac{\Sigma \sin ^2 \theta}{\rho^2} \mathrm{~d} \varphi^2,\end{align}where \begin{align}
\rho^2 & =r^2+a^2 \cos ^2 \theta, \nonumber \\
\Delta & =r^2+a^2-2 m(r) r, \nonumber \\\Sigma & =\left(r^2+a^2\right)^2-a^2 \Delta \sin ^2 \theta,\end{align}and\begin{equation}m(r)=M-\frac{Q_e^2}{2 r}+A \frac{Q_e^4}{40 r^5}.\end{equation}

We have reduced the EEH rotating black hole solution to a Kerr-Newman-like black hole one.
By setting $a = 0$, the static screened Reissner-Nordstrom solution is recovered. In order to gain some physical insight into the energy-mass function, we could allow to vary from point to point in the spacetime. In this framework, the solution behaves asymptotically as the Kerr-Newman one.

\subsection{Null Regions} 
Recall that in the Kerr case, it is well-known that if the spin parameter is extremal $(a=M)$, two horizons coincide at $r = M$. It means that the minima of the curve by the function $\Delta(r)$ coincides $r = M$ at $\Delta(r) = 0$. If $a = 0$, we expect the Schwarzschild case for the horizon where $r = 2M$. For the Reisnner-Nordstrom case, two horizons are also formed due to the effect of the black hole charge $Q_e$. Similar to the Kerr case, the critical value for the charge is $Q_e = M$. Any charge greater than $M$ would produce an imaginary horizon and unphysical. In the Kerr-Newman (KN) case, it can be shown that the critical value of charge for the horizons to be physical is $Q_e = \sqrt{M^2 - a^2}$. It implies that as $a$ tends to be large, $Q_e$ should have a small value.

\begin{figure*}
   \centering
    \includegraphics[width=0.48\textwidth]{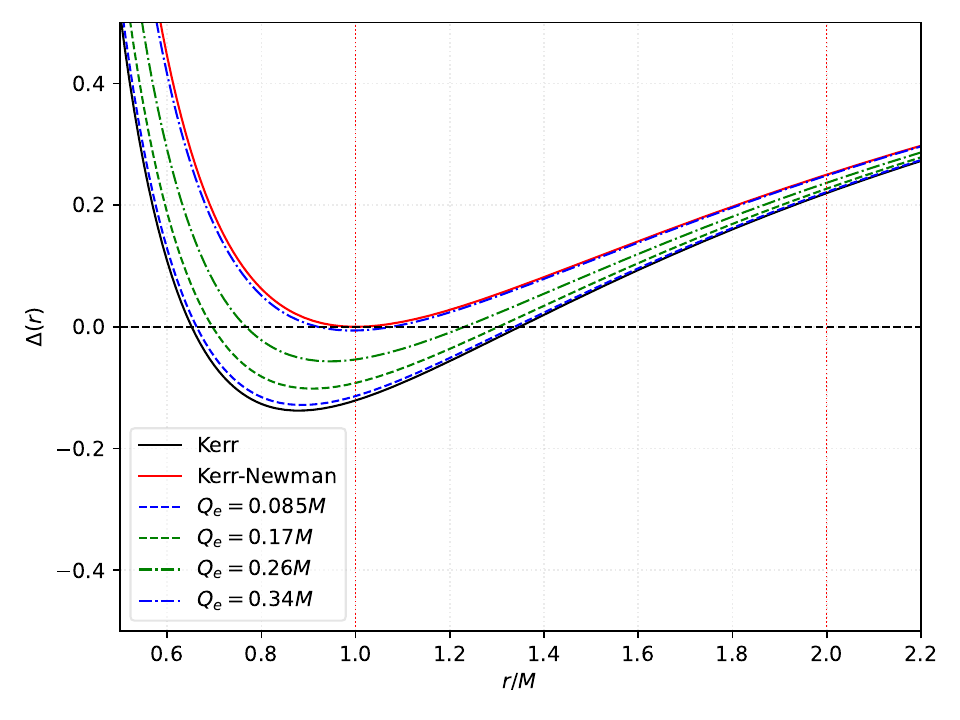}
    \includegraphics[width=0.48\textwidth]{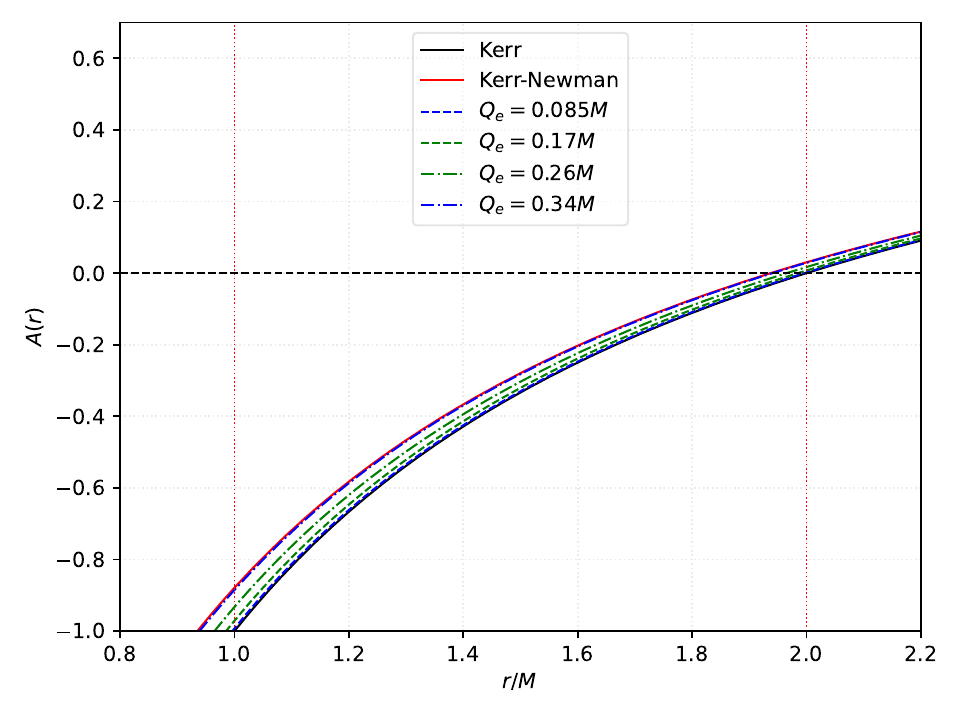}
    \caption{Top: Behavior of the event horizon. Here, we assumed $a = 0.9375M$, and $A = 1$. The highest value of charge is $Q_e \sim 0.34M$, which is the critical value of charge in the KN case. The dotted vertical line represents the horizon radius when $a = 0$ and $a = M$. Bottom: Behavior of the ergoregion when $\theta = \pi/2$.}
    \label{hor}
\end{figure*}

For the EEH black hole, we have the KN case with the addition of the screening parameter $A$ that is set to $1$. Thus, it is meaningful to explore the effect $A$ along with the spin parameter as we vary the charge $Q_e$. We numerically show this in Fig. \ref{hor}, where we include the Kerr and the KN case for comparison $(a=0.9375M)$ \cite{Cui:2023uyb}. We notice that the critical value for the horizon in the KN case is $Q_e = 0.34M$, where there are neither event nor Cauchy horizons. However, with this $Q_e$, both horizon appeared, but so close together.

Next, we examined the ergosphere radii as shown in the left panel of Fig. \ref{hor}. For the KN case, the critical value of charge must be $Q_e = \sqrt{M^2 - a^2\cos^2(\theta)}$. At $\theta = \pi/2$, this is the same as the critical value for the horizon where the minima coincide at zero. It is quite clear that $A$ has a negligible effect for the ergoregion in this case, since the dash-dot blue line coincides with that of the KN case.

\section{Null Geodesics} \label{sec3}
In this section, we start with the analysis of photon geodesic by utilizing the Hamilton-Jacobi equation, which gives  
\begin{equation} \label{e33}
    \frac{\partial S}{\partial \lambda}=-H,
\end{equation}
where $S$ is the Jacobi action, $\lambda$ is the proper time (or the affine parameter). In terms of the coordinate $x_\mu$, the Hamiltonian in General Relativity is given by
\begin{equation} \label{e34}
    H=\frac{1}{2}g^{\mu \nu }\frac{
    \partial S(\lambda)}{\partial x^{\mu }}\frac{\partial S(\lambda)}{\partial x^{\nu }},
\end{equation}
so that 
\begin{equation} \label{e35}
    \frac{\partial S}{\partial \lambda }=-\frac{1}{2}g^{\mu \nu }\frac{
    \partial S}{\partial x^{\mu }}\frac{\partial S}{\partial x^{\nu }},
\end{equation}
as follows from Eq. \eqref{e33} above. Let's use the separability ansatz for the Jacobi function
\begin{equation} \label{e36}
    S=\frac{1}{2}\mu ^{2}\lambda -Et+L\phi +S_{r}(r)+S_{\theta }(\theta),
\end{equation}
and with the particle mass $\mu$, one can obtain the following first-order motion equations \cite{Slany:2020jhs}
\begin{align} \label{eos1}
    &\Sigma\frac{dt}{d\lambda}=\frac{\Xi (r^2+a^2)P(r)}{\Delta _r}-\frac{\Xi aP(\theta )}{\Delta _{\theta }}, \nonumber \\
    &\Sigma\frac{dr}{d\lambda}=\sqrt{R(r)}, \nonumber \\
    &\Sigma\frac{d\theta}{d\lambda}=\sqrt{\Theta(\theta)}, \nonumber \\
    &\Sigma\frac{d\phi}{d\lambda}=\frac{\Xi aP(r)}{\Delta _r}-\frac{\Xi P(\theta )}{\Delta _{\theta }\sin ^2\theta },
\end{align}
where 
\begin{align} \label{eos}
    &R(r) = P(r)^2 - \Delta _r(\mu^2r^2+K), \nonumber \\
    &P(r) = \Xi E(r^2+a^2)-\Xi aL, \nonumber \\
    &\Theta(\theta) = \Delta_{\theta }(K-\mu^2a^2\cos ^2\theta ) - \frac{P(\theta )^2}{\sin^2\theta }, \nonumber \\
    & P(\theta) = \Xi(aE\sin ^2\theta -L ).
\end{align}
The consequence of a hidden symmetry in the $\theta$-coordinate \cite{Slany:2020jhs,Carter:1968rr} gives a constant of motion $K = \Xi^2(aE-L)^2$, found in the third equation in Eq. \eqref{eos} above

The geodesic of massless particles can be easily studied by setting $\mu = 0$. In determining the unstable circular orbit of photons, the condition below must be satisfied:
\begin{equation} \label{e39}
    R(r)=\frac{dR(r)}{dr}\Big|_{r=r_{\rm ps}} =0.
\end{equation}
The photon-sphere region is deeply related to the shadow cast by a black hole. A small perturbation on the orbit may cause photons to escape from $r_{\rm ps}$ to infinity, and then reach the observer's detectors. This is called backward ray tracing. For photons, it is always useful to define two impact parameters:
\begin{equation} \label{e45}
    \xi=\frac{L}{E} \quad {\rm and}\quad \eta=\frac{K}{E^2}.
\end{equation}
The former is the impact parameter associated to the $\phi$ coordinate, while the latter is to the $\theta$ coordinate, which is a generalization to include latitudinal motion contribution of photons. Using the function $R(r)$ in Eq. \eqref{eos} and the condition present in Eq. \eqref{e39}, the above quantities are given explicitly by
\begin{align} 
    \xi&=\frac{\Delta_r'(r^{2}+a^{2})-4\Delta_r r}{a\Delta_r'},\nonumber \\
    \eta&=\frac{-r^{4}\Delta_r'^{2}+8r^{3}\Delta_r\Delta_r'+16r^{2}\Delta_r(a^{2}-\Delta_r)}{a^{2}\Delta_r'^{2}}, \label{e47}
\end{align}
which is a convenient expression due to the fact that it can describe any black hole model described by the function $\Delta_r$. 
Depending on how complicated $\Delta_r$ is, analytic or numerical values of $r_{\rm ps}$ can be sought off by solving $r$ in $\eta = 0$. The analytical solutions are well-known for both Schwarzschild and Kerr black holes (which have two values for $r_{\rm ps}$).
\begin{figure*}
   \centering
    \includegraphics[width=0.48\textwidth]{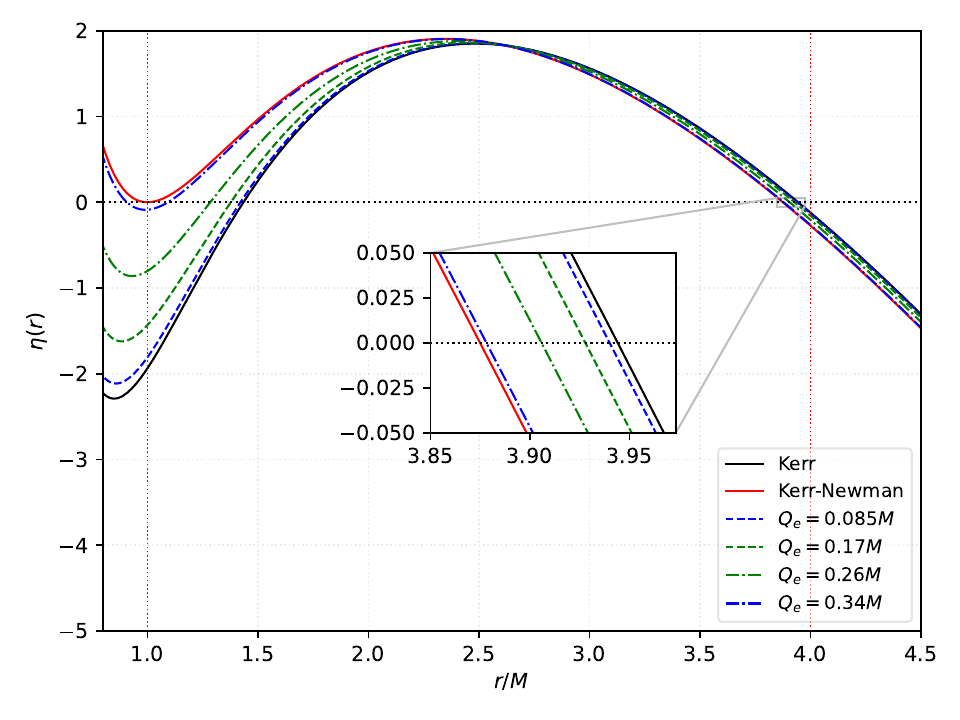} 
    \includegraphics[width=0.48\textwidth]{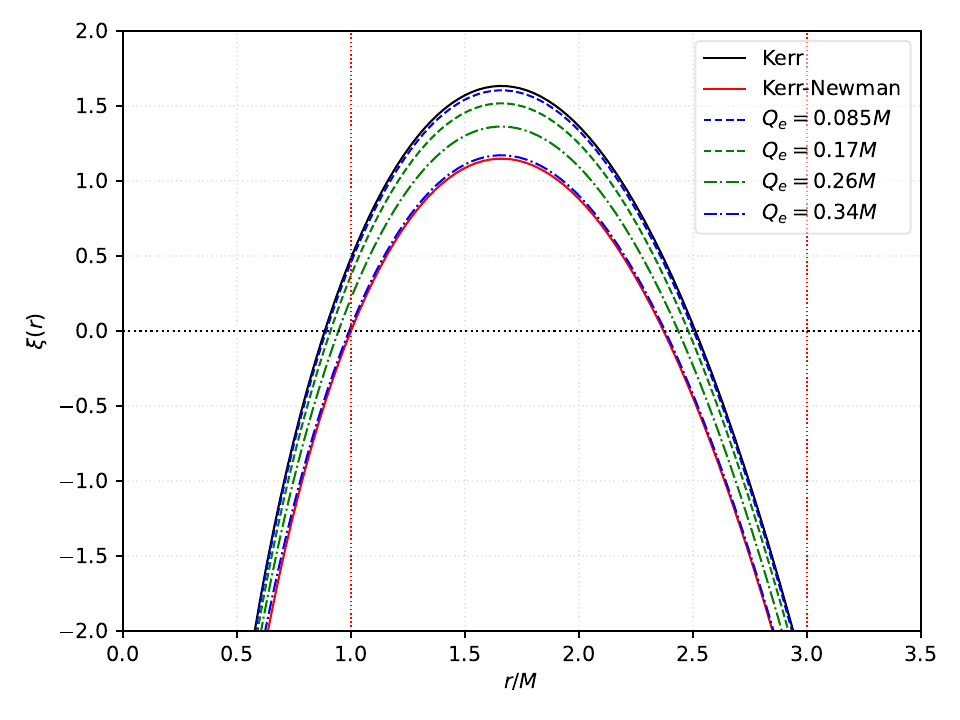}
    \caption{Left panel: retrograde and prograde photons. Right panel: zero angular momentum orbit. Here, $a = 0.9375M$ as we vary $Q_e$ and setting $A = 1$.}
    \label{rph}
\end{figure*}
The plot is shown in Fig. \ref{rph} for three cases: retrograde, prograde, and the orbit that defines zero angular momentum. For a given spin parameter, we see in the inset plot on the left panel how the black hole charge $Q_e$ causes deviation from the KN case (that is, to increase the value of the retrograde orbit slightly). A more noticeable increase is shown for the prograde orbit.

\section{Shadow cast and observables} \label{sec4}
\begin{figure*}
   \centering
    \includegraphics[width=0.48\textwidth]{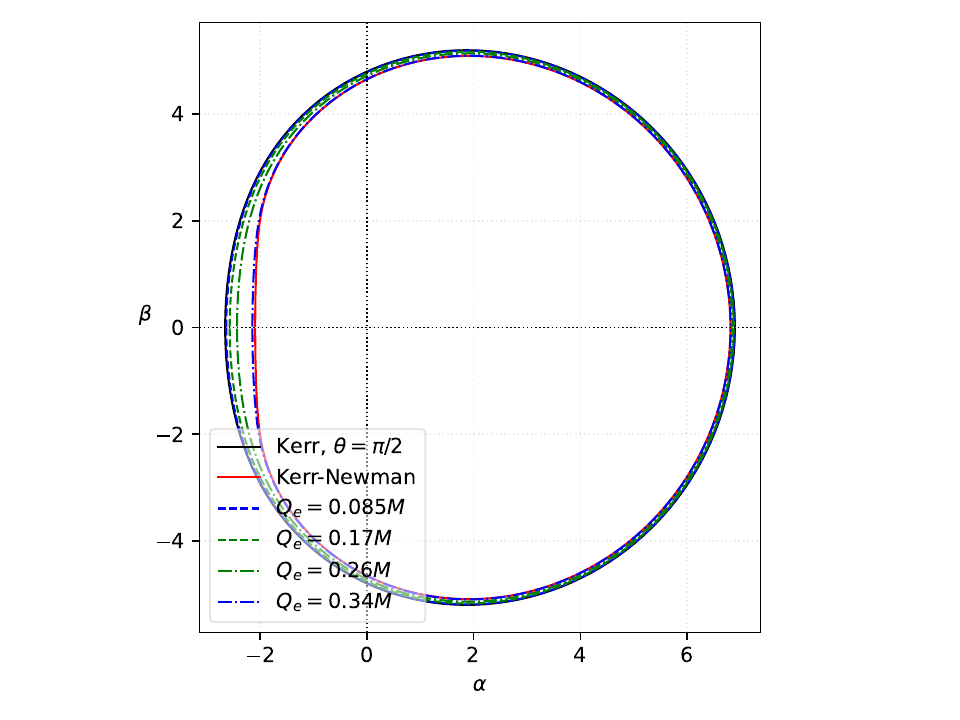} 
    \includegraphics[width=0.48\textwidth]{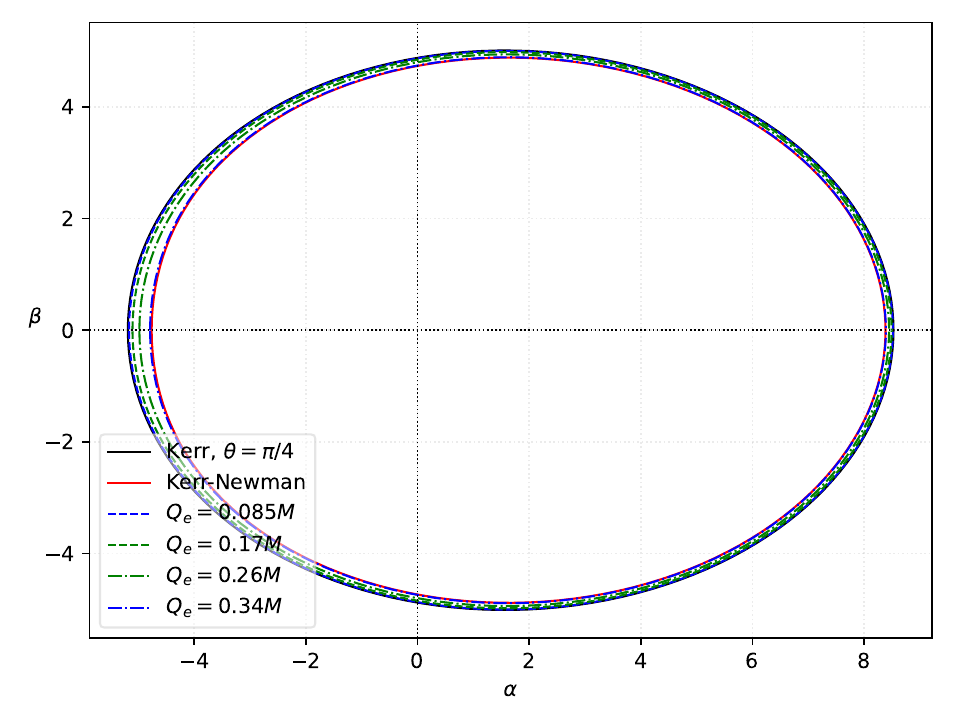}
    \caption{Left panel: Case of different values of $Q_e$ at inclination $\theta = \pi/2$, $A = 1$, and $a = 0.9375M$. Right panel: For $\theta = \pi/4$. The Kerr and Kerr-Newman cases are included for comparison.}
    \label{sha_cast}
\end{figure*}

As mentioned earlier, escaping photons define the shadow cast and it can be done by using the celestial coordinates of the observer at $(r_{\rm o},\theta_{\rm o})$. Such an observer is also known as the Zero Angular Momentum Observer (ZAMO). The general definition of the celestial coordinates is \cite{Johannsen:2013vgc}
\begin{align} \label{e48}
    \alpha  &=-r_{\rm o}\frac{\xi }{\zeta\sqrt{g_{\phi \phi }} \left( 1+\frac{g_{t\phi }}{g_{\phi \phi }}\xi\right) }, \nonumber\\
    \beta  &= r_{\rm o}\frac{\pm \sqrt{\Theta (\theta_o)}}{\zeta\sqrt{g_{\theta \theta }} \left( 1+\frac{g_{t\phi}}{g_{\phi \phi }}\xi \right) },
\end{align}
and the condition $r_{\rm o} \rightarrow \infty$  leads to the simplified relations
\begin{align} \label{e49}
    \alpha&=-\xi \csc \theta_{\rm o},   \nonumber \\
    \beta&=\pm \sqrt{\eta +a^{2}\cos ^{2}\theta_{\rm o}-\xi ^{2}\cot^{2}\theta_{\rm o}}.
\end{align}
If the observer is at the equatorial plane $\theta_{\rm o}=\pi/2$, these expressions reduce to $\alpha = 0$ and $\beta= \pm \sqrt{\eta}$. Furthermore, when $a=0$, we obtain the shadow cast as a perfect circle. The plot of $\beta$ vs. $\alpha$ is shown in Fig. \ref{sha_cast} for the black hole spin parameter value of  $a=0.9375M$. We added the Kerr and the KN cases for comparison.

Due to the high spin that we considered, the D-shaped nature of the shadow cast manifests clearly for the Kerr case. Relative to this, the D-shape is more pronounced for the KN case as we add the charge $Q_e$. The prograde orbit pronounced the effect of $A$ relative to the KN case, and it confirms the observer in Fig. \ref{rph}.

\begin{figure*}
   \centering
    \includegraphics[width=0.48\textwidth]{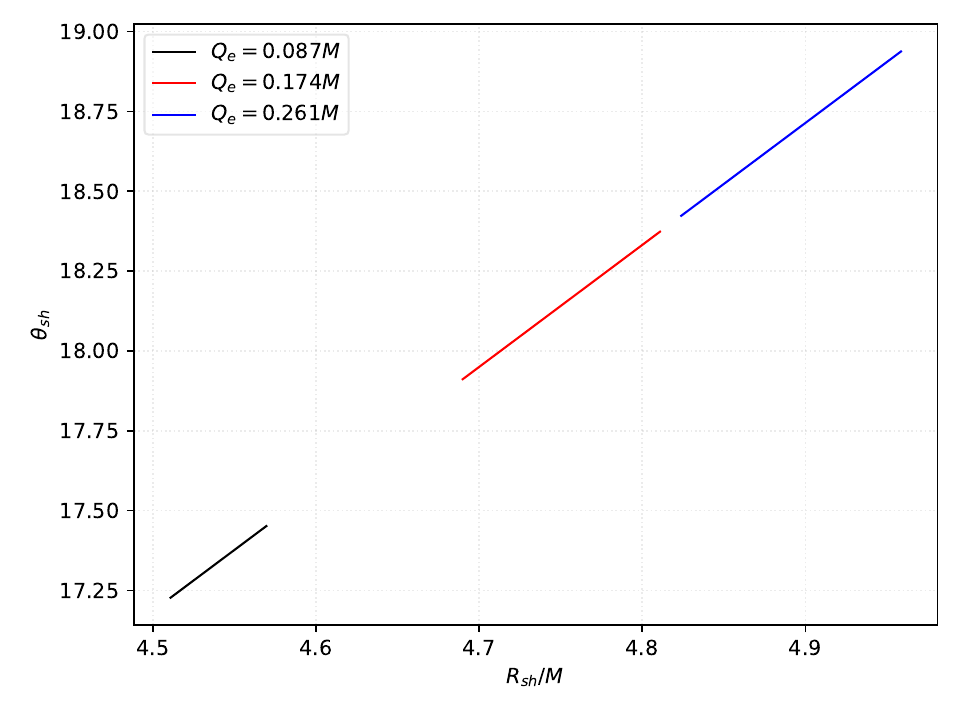} 
    \includegraphics[width=0.48\textwidth]{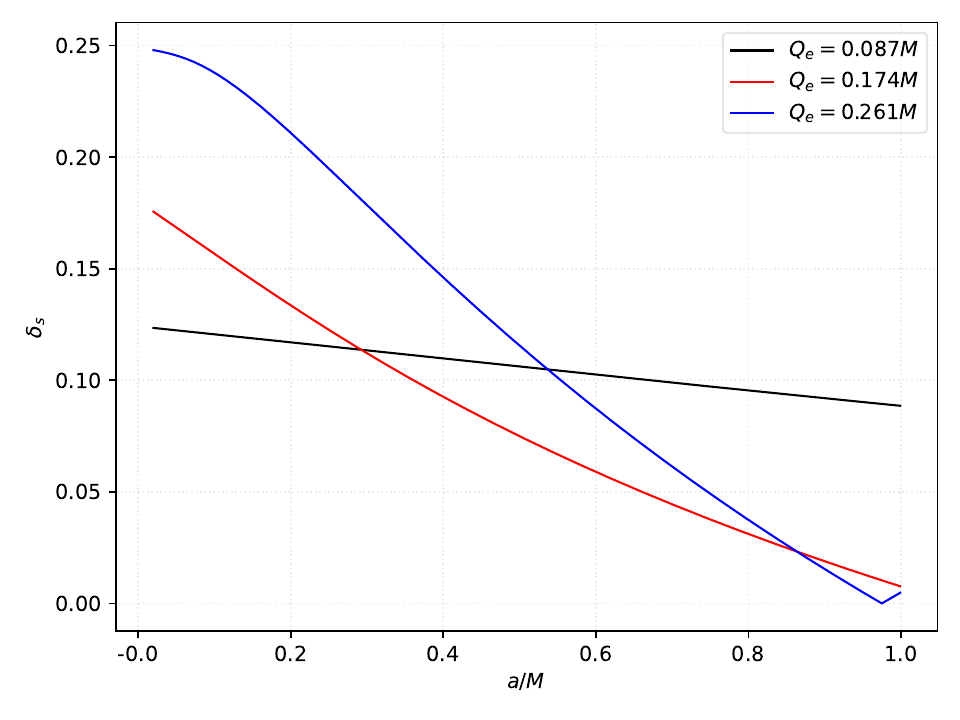}
    \includegraphics[width=0.48\textwidth]{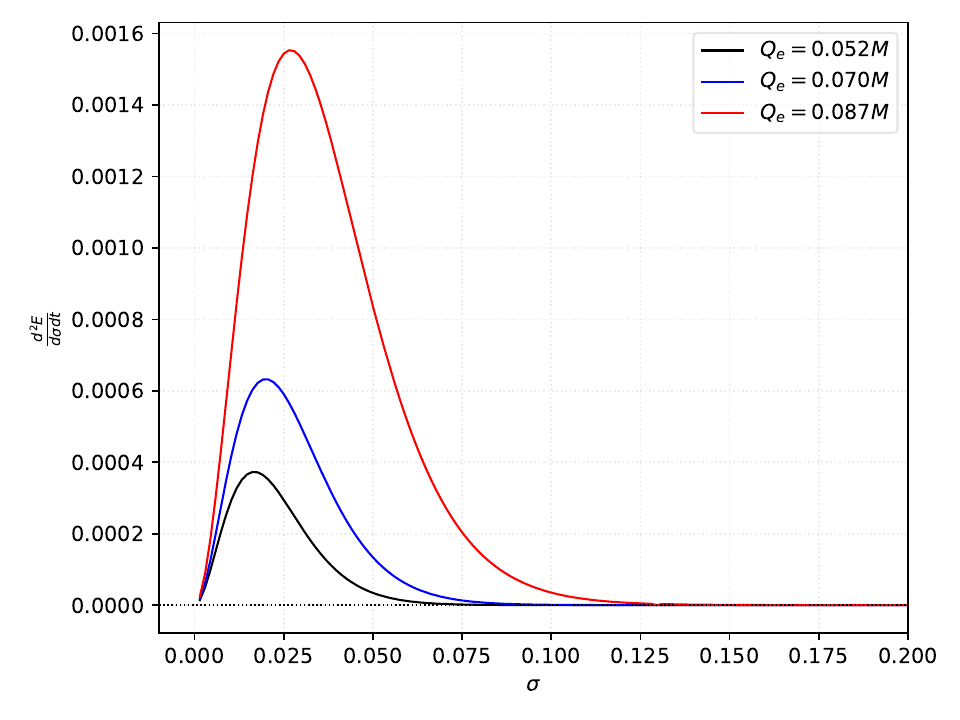} 
    \caption{Upper left panel: Angular radius using the mass of M87* and our actual distance. Upper right panel: Behavior of the distortion parameter as $A=1$ for different $Q_e$. Lower panel: The energy emission rate as frequency $\sigma$ varies for different $Q_e$ (Note that this is also for $A=1$).}
    \label{shaobs}
\end{figure*}

As the spin parameter $a$ becomes more extremal, one can still obtain the shadow radius defined by $R_{\rm sh}$. Its numerical value can be calculated via \cite{Hioki:2009na,Dymnikova:2019vuz}
\begin{equation} \label{eq-rad}
    R_{\rm sh}=\frac{\beta_{\text{t}}^2+(\alpha_{\text{t}}-\alpha_{\text{r}})^2}{2|\alpha_{\text{t}}-\alpha_{\text{r}}|}.
\end{equation}
 \textcolor{black}{Using the above equation, we can define the shadow's angular radius $\theta_{\rm sh}$:}
\begin{equation} \label{eq-ang}
    \theta_{\rm sh}=9.87098\times10^{-3} \frac{R_{\rm sh}M}{D},
\end{equation}
where $M$ is the black hole's mass in units of $M_{\odot}$, and $D$ is measured in parsec. \textcolor{black}{ We plot the numerical result in Fig. \ref{shaobs} at the upper left panel, which is consistent with Fig. \ref{sha_cast}. Other observables that can be derived from the shadow are the distortion parameter $\delta_{\rm sh}$ and the energy emission rate $\frac{d^{2}E}{d\omega dt}$, which are defined as follows:}
\begin{equation} \label{eq-dis}
    \delta_{\rm sh}=\frac{d_{\rm sh}}{R_{\rm sh}}=\frac{\tilde{\alpha}_{\text{l}}-\alpha_{\text{l}}}{R_{\rm sh}},
\end{equation}
\begin{equation} \label{eq-eer}
    \frac{d^{2}E}{d\omega dt}=2\pi^{2}\frac{\Pi_{ilm}}{e^{\omega/T}-1}\omega^{3}.
\end{equation}
We can approximate the energy absorption cross-section as $\Pi_{ilm} \sim \pi R_{\rm sh}^2$ for an observer at $r_{\rm o} \rightarrow \infty$. We plot these two observables in Fig. \ref{shaobs} upper right panel and lower panel, where we can see how these observables behave due to the effect of the black hole charge $Q_e$. As for the energy emission rate, as it is related to the black hole's lifetime, higher $Q_e$ makes the EEH black hole emit more energy. Also, we observe that peak frequency shifts to higher values as $Q_e$ increases.

\section{QNMs using WKB approximation} \label{sec5}
This section deals with the QNMs of both the rotating and static black holes by using WKB approximation methods. QNMs, which characterize the damped oscillations of perturbations in the black hole spacetime, provide critical insights into the stability and properties of these celestial objects. The WKB (Wentzel - Kramers -Brillouin) approximation method, known for its effectiveness in semi-classical analyses, is employed to derive and analyze the frequencies and damping rates of these oscillations. By applying this method, we aim to elucidate the differences and similarities in the QNM spectra of rotating versus static black holes, thereby enhancing our understanding of their dynamic behaviours under perturbations.

\subsection{ The rotating case}
In this subsection, we shall derive equations representing the frequencies of QNMs associated with this rotating black hole using the WKB approximation. This requires determining both the real and imaginary components of $\omega$ with leading and next-to-leading order accuracy. Our focus will be on obtaining an analytical approximation for the frequency spectrum corresponding to this black hole solution. Although higher-order WKB methods are more reliable, in the case of rotating black holes, due to the complexity of the method, we shall use the WKB method only up to the leading order corrections.

Before we delve into the specifics of our results pertaining to the angular and radial Teukolsky equations, it is essential to revisit a fundamental aspect of the WKB expansion. This concept will be a recurring theme throughout our paper. For a more comprehensive discussion of WKB methods, please refer to \cite{Iyer:1986np, Dias:2022oqm, Konoplya:2019hlu, Konoplya:2011qq,Konoplya:2017wot,Konoplya:2003ii,Yang:2012he}.

In the beginning, we consider a wave equation for the wave function $\psi(x)$, given by:

\begin{equation}
\epsilon^2\frac{d^2 \psi}{d x^2} + U(x) \psi = 0.
\end{equation}
In the above equation, $\epsilon$ is a small positive number. For this equation, the solution can be expended in the following form:
\begin{equation}
\psi = e^{\frac{\mathcal{S}_0}{\epsilon} + \mathcal{S}_1 + \epsilon S_2 + \ldots}.
\end{equation}
In the above expansion, the primary and secondary variables {\it i.e.,} $\mathcal{S}_0$ and $\mathcal{S}_1$ can be expressed in the following form \cite{Yang:2012he}:
\begin{subequations}
\begin{align}
\mathcal{S}_0 &= \pm i \int ^x \sqrt{U(x)}\, dx, \\
\mathcal{S}_1 &=-\frac{1}{4}\log{U(x}).
\end{align}
\end{subequations}
These formulas will serve as the foundation for our examination of the radial and angular Teukolsky equations in the subsequent parts of our study.

Teukolsky demonstrated that the Kerr spacetime's scalar, vector, and tensor perturbations all adhere to a unified master equation for scalar variables with spin weight $\bar{s}$. Furthermore, this master equation can be solved through a separation of variables approach \cite{Teukolsky:1972my}. We will employ the variable $u$ to represent our scalar field, and we shall decompose this scalar wave as follows \cite{Luna:2022rql, Yang:2012he, Yang:2021zqy}:
\begin{equation}
\label{eq6}
u(t,r,\theta,\phi) = e^{-i \omega t }e^{i m_l \phi}u_r(r)u_{\theta}(\theta).
\end{equation}
Subsequently, at the relevant order for $l \gg 1$, the angular equation for $u_\theta(\theta)$ can be expressed as:
\begin{align}
\label{AngTeuk:theta}
\frac{1}{\sin \theta}\frac{d}{d\theta}\left[\sin{\theta} \frac{d u_{\theta}}{d \theta}\right] +
\left[a^2\omega^2\cos^2{\theta}-\frac{m_l^2}{\sin^2{\theta}}+\mathcal{A}_{lm_l}\right] u_{\theta}=0,
\end{align}
where $\mathcal{A}_{lm_l}$ represents the angular eigenvalue of this equation. The equation governing the radial function $u_r(r)$ is given by:

\begin{equation}
\label{eqr}
\frac{d^2 u_r}{d r_*^2} + \frac{K^2-\Delta\lambda^0_{l m_l}}{(r^2+a^2)^2} u_r=0,
\end{equation}
with the definitions of the parameters as follows:
\begin{align}
K&=-\omega(r^2+a^2)+a m_l, \\
\lambda^0_{l m_l} &= \mathcal{A}_{lm_l}+a^2\omega^2-2a m_l \omega.
\label{eqexplan}
\end{align}
It is worth noting that in our calculations, we have neglected higher-order terms based on the facts that $\omega_R \sim O(l)$, $\omega_I \sim O(1)$, and $m_l \sim O(l)$ in comparison to the terms we are considering. Therefore, the spin parameter associated with the perturbation has no impact on the equations governing QNMs of the black hole spacetime. In the above equation governing the behaviour of the radial function $u_r$, the rate of variation is calculated with respect to the tortoise coordinate $r_*$ which is defined as,
\begin{equation*}
     \frac{d}{dr_*} \equiv \frac{\Delta}{r^2+a^2}\frac{d}{dr}
\end{equation*}

We can derive an expression for $\mathcal{A}_{lm_l}$ in terms of $\omega$, $l$, and $m_l$ by analyzing the angular equation within the WKB approximation. To begin, let us outline our approach to this calculation. Given that the frequency $\omega = \omega_R - i \omega_I$ is complex, the angular eigenvalue $\mathcal{A}_{lm_l}$, which depends on $\omega$, must also be complex \cite{Yang:2012he, Yang:2021zqy}. We express this as:
\begin{equation}
\mathcal{A}_{lm_l} = A^R_{lm} + i A^I_{lm},
\end{equation}
to distinguish between the real and imaginary components.
By utilizing perturbation theory for eigenvalue equations, we find:
\begin{equation}
\mathcal{A}_{lm_l}^I = -2a^2\omega_R\omega_I \langle\cos^2\theta\rangle,
\label{Alimg}
\end{equation}
where the expectation value is given by the following expression:
\begin{align}
\langle\cos^2\theta\rangle &= \frac{\displaystyle\int \cos^2\theta |u_\theta|^2\sin\theta d\theta}{\displaystyle\int |u_\theta|^2\sin\theta d\theta}  \nonumber\\
&=\frac{\displaystyle \int_{\theta_-}^{\theta_+} \frac{\cos^2\theta}{\sqrt{a^2\omega_{R}^2 \cos^2\theta -\frac{m_l^2}{\sin^2\theta}+\mathcal{A}_{lm_l}^R}}  d\theta}{\displaystyle \int_{\theta_-}^{\theta_+} \frac{1}{\sqrt{a^2\omega_{R}^2 \cos^2\theta -\frac{m_l^2}{\sin^2\theta}+\mathcal{A}_{lm_l}^R}} d\theta}.
\end{align}

The Bohr-Sommerfeld condition for such a case from Ref. \cite{Yang:2012he} can be given as
\begin{equation}
\label{BS_eq}
\int_{\theta_-}^{\theta_+}
d\theta \sqrt{a^2\omega_{R}^2 \cos^2\theta -\frac{m^2}{\sin^2\theta}+\mathcal{A}_{lm}^R} 
=\left(l+1/2-|m|\right)\pi\,.
\end{equation}

By differentiating the Bohr-Sommerfeld condition \eqref{BS_eq} with respect to the variable $z = a \omega_R$ and considering the parameter $\mathcal{A}_{lm_l}$ as a function of $z$, we can reformulate the expression as follows:
\begin{align}
\langle\cos^2\theta\rangle &= \left. - \frac{1}{2z}\frac{\partial \mathcal{A}_{lm_l}^R(z)}{\partial z}\right|_{z=a \omega_R }.
\end{align}
We use this expression into the Eq. \eqref{Alimg} to obtain the following relation
\begin{equation} \label{almI}
\mathcal{A}_{lm_l}^I = a \omega_I \left[\frac{\partial \mathcal{A}_{lm_l}^R(z)}{\partial z}\right]_{z=a \omega_R}.
\end{equation}

This Eq. \eqref{almI} outlines a numerical method for determining $\mathcal{A}_{lm_l} = \mathcal{A}_{lm_l}^R + i \mathcal{A}_{lm_l}^I$ for a rotating black hole. As expected, the term is a complex quantity. The real part of it is associated with the oscillation frequency of ring-down GWs. An approximation of this relation gives us,
\begin{equation}
\mathcal{A}_{lm_l} \approx (l+1/2)^2 - \frac{a^2\omega^2}{2}\left[1-\frac{m_l^2}{(l+1/2)^2}\right].
\end{equation}
After calculating the angular eigenvalues $\mathcal{A}_{lm_l}$ in terms of the oscillation frequency $\omega$, we turn our focus to the radial Teukolsky equation. As shown in Eq.~\eqref{eqr}, the radial equation is formulated as:
\begin{equation}
\frac{d^2 u_r}{dr_*^2} + V^r u_r = 0,
\label{schRad}
\end{equation}
where the potential associated with the above equation can be expressed as
\begin{equation}
V^r(r,\omega) = \frac{[\omega(r^2+a^2)-m_l a]^2 -\Delta\left[\mathcal{A}_{lm_l}(a\omega) +a^2\omega^2 -2m_l a\omega\right]}{(r^2+a^2)^2}.
\end{equation}
Following Ref. \cite{Yang:2012he}, the leading-order WKB approximant for $u_r$ can be expressed as:
\begin{equation}
\label{WKBradial}
u_r = b_+ e^{i\int^{r_*} \sqrt{V^r(r_*')} dr_*'} + b_- e^{-i\int^{r_*} \sqrt{V^r(r_*')} dr_*'}.
\end{equation}
The outgoing mode $r_*\rightarrow +\infty$ and the ongoing mode $r_*\rightarrow -\infty$ demands that
\begin{equation}
u_r = b_+ e^{i\int^{r_*} \sqrt{V^r(r_*')} dr_*'}
\end{equation}
for the region having $r\rightarrow +\infty$, and
\begin{equation}
u_r = b_- e^{-i\int^{r_*} \sqrt{V^r(r_*')} dr_*'}
\end{equation}
for the region having $r_*\rightarrow -\infty$. 
In simpler terms, a solution to Eq.\ (\ref{schRad}) will exhibit the specified asymptotic behavior if $ V^r \approx 0$ at a point  $r = r_0$, with  $V^r$  being positive on both sides of this point. This allows the WKB expansion \eqref{WKBradial} to be applied in the regions flanking $r = r_0$. However, the solution near $r_0$ must be determined separately and matched with the WKB approximation to constrain the frequency and thereby determine $\omega$ \cite{Yang:2012he}. Iyer and Will performed an extensive calculation of this procedure to high orders in the WKB approximation. The main difference between their calculation and ours at lower orders is due to the more complex dependence of $V^r$ on $\omega$ in our case, especially because $\mathcal{A}_{lm_l}$ depends on $\omega$ in a more intricate manner. As stated in Ref. \cite{Iyer:1986np}, the conditions that need to be solved at the leading and next-to-leading orders to determine $\omega_R$ are:
\begin{equation}
\label{Vreq}
V^r(r_0,\omega_R) = \left.\frac{\partial V^r}{\partial r}\right|_{(r_0,\omega_R)} = 0.
\end{equation}
Further, it is possible to write these conditions as:
\begin{align}\label{v}
\Omega_R &= \frac{ \mu a}{r_0^2+a^2} \pm \frac{\sqrt{\Delta(r_0)}}{r_0^2+a^2}\beta(a \Omega_R) \,, \\
0 &=\frac{\partial}{\partial r}\left[\frac{\Omega_R (r^2+a^2)- \mu a}{\sqrt{\Delta(r)}} \right]_{r=r_0} \,, \label{dvdrG}
\end{align}
where $\Omega_R = (l+1/2) \omega_R$ and $\beta(a \Omega_R) = \sqrt{\frac{1}{2} a \Omega _R \left(a \left(\mu ^2+1\right) \Omega _R-4 \mu \right)+1}$.
From (\ref{dvdrG}) condition, one gets
\begin{equation}
    \omega_R = \frac{a m_l \left(A Q_e^4+10 r_0^5 \left(r_0-M\right)\right)}{ \left(a^2 \left(A Q_e^4-10 r_0^5 \left(M+r_0\right)\right)+2 r_0^2 \left(A Q_e^4-5 r_0^4 \left(-3 M r_0+2 Q_e^2+r_0^2\right)\right)\right)}
\end{equation}
The imaginary part $\omega_I$ can be calculated in the leading order using the method described by Iyer and Will in their work \cite{Iyer:1986np}. This approach yields the result that:
\begin{align}
\omega_I & =-(n+1/2) \frac{\sqrt{2\left(\frac{d^2 V^r}{dr_*^2}\right)_{r_0,\omega_R}}}{\left(\frac{\partial V^r}{\partial \omega}\right)_{r_0,\omega_R}}.
\end{align}

This expression shows that the leading order imaginary part $\omega_I$ depends on the overtone number $n$ and is proportional to the square root of the second derivative of the potential with respect to the tortoise coordinate. Although this relation does not provide imaginary QNMs or damping rate of ring-down GWs up to the desired accuracy to compare with the observational results, it is still possible to theoretically understand the behavior of the ring-down modes by using this relation. In the case of rotating black holes, the higher-order corrections become too complex and hence we have limited our investigation to the leading orders only.
For our case, the above relation, under a suitable approximation of the angular eigenvalues, gives

\begin{equation}
    \omega_I = -\frac{(2 l+1)^3 (2 n+1) \left(20 r_0^4 \left(a^2-2 M r_0+Q_e^2+r_0^2\right)-A Q_e^4\right) \sqrt{\frac{8 \omega _R^2 \left(a^2 \left(12 l (l+1)-4 m_l^2+3\right)+12 (2 l+1)^2 r_0^2\right)}{(2 l+1)^4}-4}}{4 \omega _R \left(C_1+20 r_0^4 C_2\right)-8 a (2 l+1)^2 m_l \left(A Q_e^4+40 M r_0^5-20 Q_e^2 r_0^4\right)},
\end{equation}
where 
\begin{eqnarray}
C_1 &=& a^2 A Q_e^4 \left((2 l+1)^2+4 m_l^2\right)\,, \\
C_2 &=& a^4 \left((2 l+1)^2-4 m_l^2\right)+2 a^2 M r_0 \left((2 l+1)^2+4 m_l^2\right)-a^2 Q_e^2 [(2 l+1)^2+4 m_l^2]+a^2 r_0^2 \left(12 l (l+1)
     - 4 m_l^2+3\right) +  \nonumber \\
    &&     +2 (2 l+1)^2 r_0^4.
\end{eqnarray}

\begin{figure*}[t!]
      	\centering{
      	\includegraphics[scale=0.55]{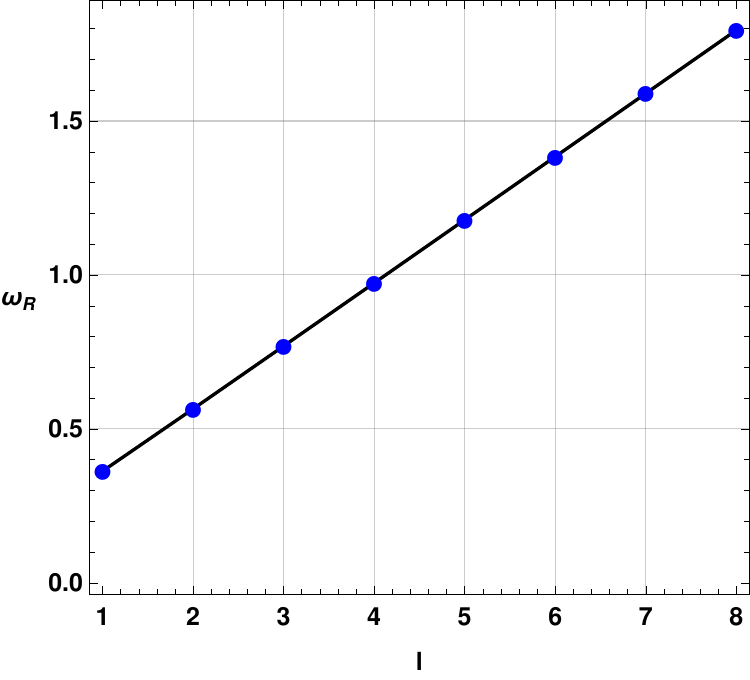}
       \includegraphics[scale=0.55]{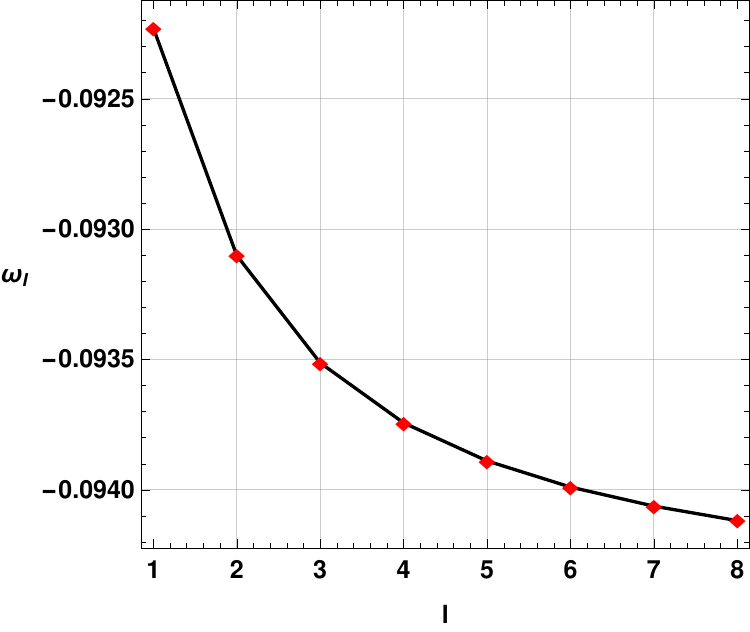}}
      	\caption{Variation of real and imaginary QNMs using $M = 1, Q_e = 0.5, A = 1, a = 0.5, n = 0$ and $ m_l = 1$. }
      	\label{figQNM01}
      \end{figure*}

      \begin{figure*}[t!]
      	\centering{
      	\includegraphics[scale=0.55]{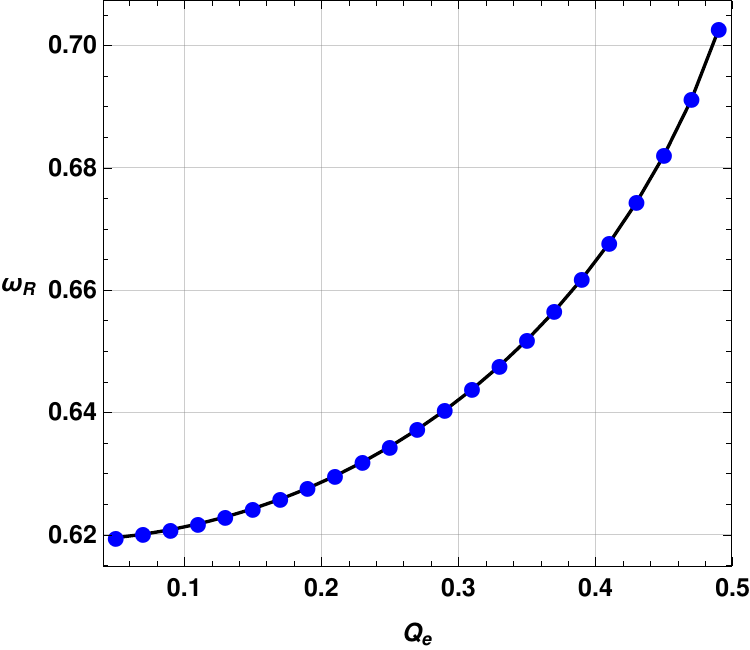}
       \includegraphics[scale=0.56]{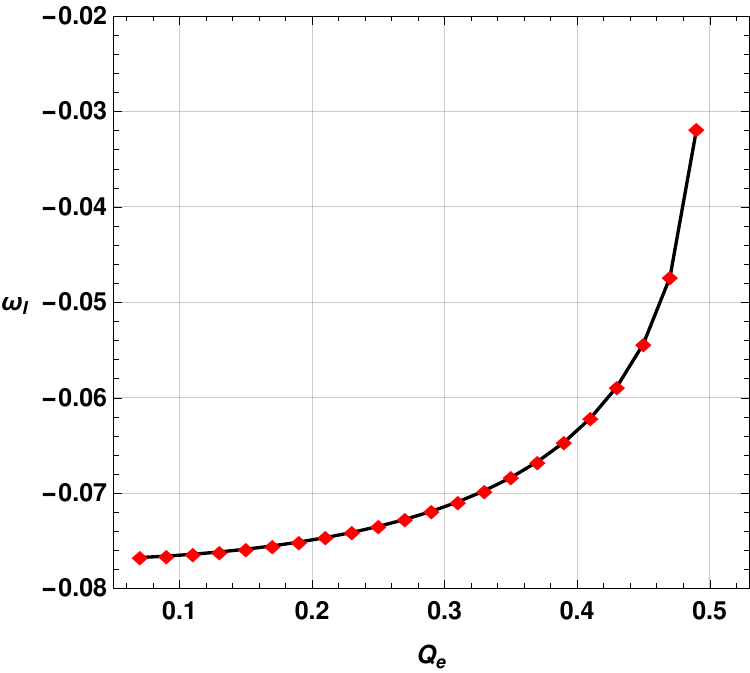}}
      	\caption{Variation of real and imaginary QNMs using $M = 1, A = 1, a = 0.9, n = 0, l = 2$ and $ m_l = 1$.}
      	\label{figQNM02}
      \end{figure*}

      \begin{figure*}[t!]
      	\centering{
      	\includegraphics[scale=0.55]{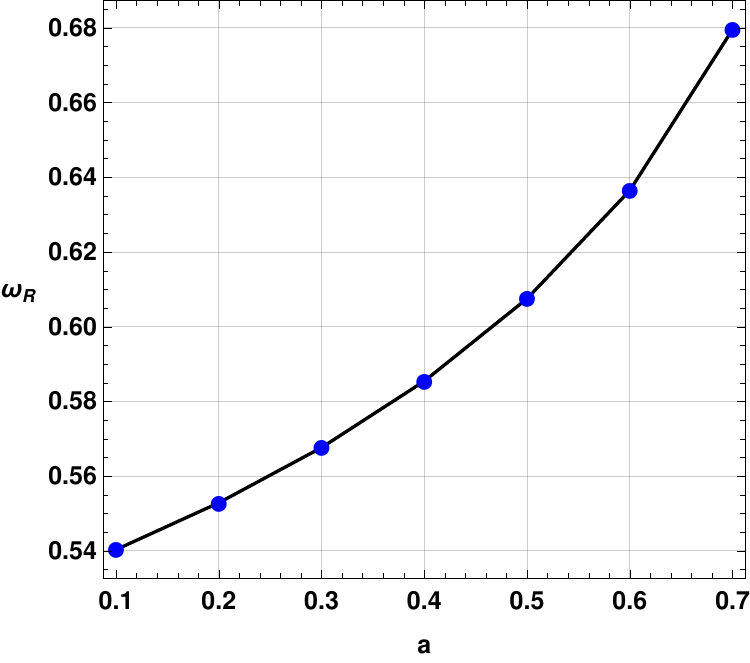}
       \includegraphics[scale=0.56]{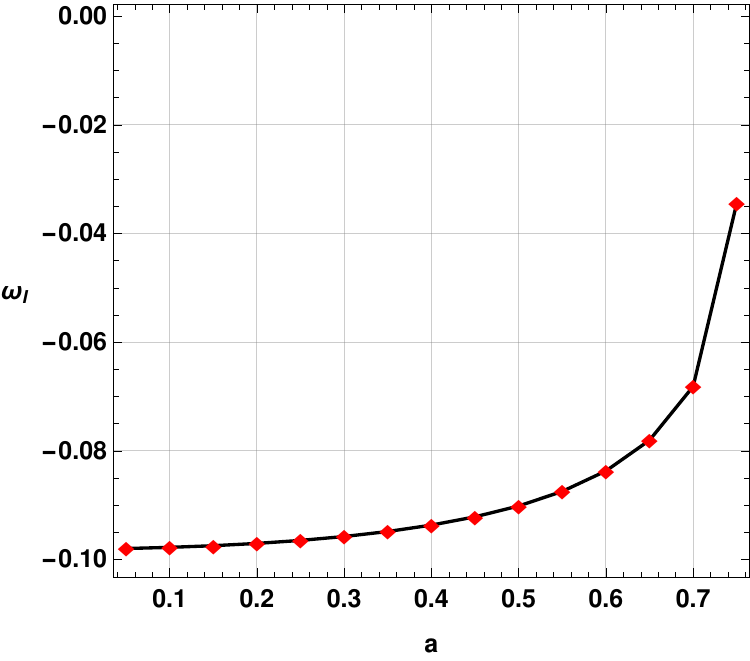}}
      	\caption{Variation of real and imaginary QNMs using $M = 1, Q_e = 0.7, A = 1, n = 0, l =2$ and $ m_l = 1$.}
      	\label{figQNM04}
      \end{figure*}
      
We have plotted the variation of QNMs for different values of multipole moments $l$ in Fig. \ref{figQNM01} using the above relations. In Fig. \ref{figQNM02}, we have shown how the QNMs vary with respect to the charge of the black hole $Q_e$. One may note that the real QNMs increase non-linearly with the value of $Q_e$. On the other hand, the damping rate of GWs decreases non-linearly with an increase in $Q_e$. Finally, from Fig. \ref{figQNM04}, it is clear that for a rotating black hole, the oscillation frequency of ring-down GWs increases, and the damping rate decreases with an increase in the value of $a$. The variation of damping rate or decay rate of ring-down GWs, as seen from the figure, is non-linear with respect to the parameter $a$. However, if we compare it with the Fig. \ref{figQNM02}, the charge parameter $Q_e$ has a more significant impact on the damping rate than that of the parameter $a$.

\subsection{Static case}
\begin{figure*}[t!]
      	\centering{
      	\includegraphics[scale=0.75]{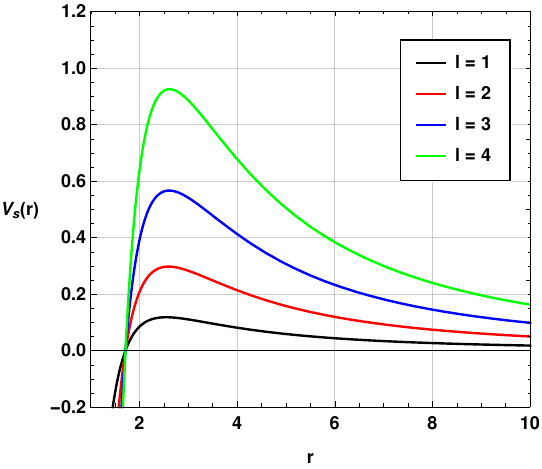}
       \includegraphics[scale=0.75]{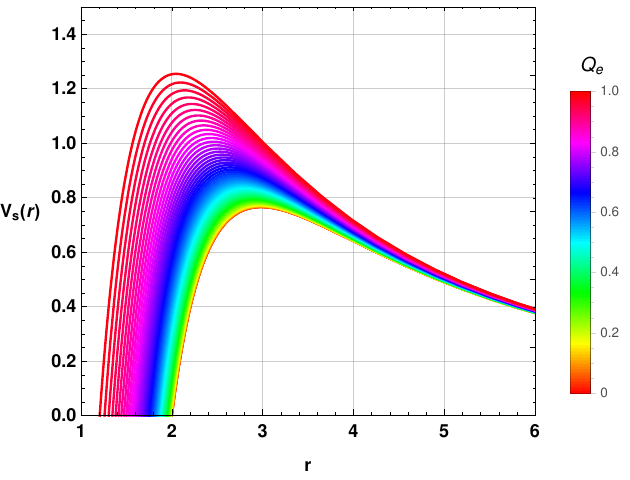}}
      	\caption{Variation of scalar potential using $M = 1$ and $a=0$. On the first panel, $Q_e = 0.7$, $A = 1$, and on the second panel, $l = 4$ and $A = 1$ have been used. }
      	\label{static01}
      \end{figure*}

In this part, we have calculated the QNMs of the black hole using $a=0$ {\it i.e.}, static case. In this scenario, the scalar potential associated with the black hole becomes:
\begin{equation}
    V_s (r) = \left(1-\frac{2 M}{r}+\frac{Q_e^2}{r^2}-\frac{A Q_e^4}{20 r^6}\right) \left(\frac{3 A Q_e^4}{10 r^8}+\frac{2 M}{r^3}-\frac{2 Q_e^2}{r^4}+\frac{l (l+1)}{r^2}\right).
\end{equation}
We have shown the variation of the potential in Fig. \ref{static01} for different values of multipole moment $l$, and black hole charge $Q_e$. With an increase in the charge parameter $Q_e$, the peak value of the potential increases and shifts slightly towards the event horizon of the black hole. Since the potential behaviour depends significantly on the model parameters, it suggests that the model parameters may have noticeable impacts on the QNMs spectrum of the static black hole. Moreover, the WKB method deals with the maximum potential and from this perspective, it seems that the model parameter $Q_e$ might have noticeable impacts on the ring-down GWs.

      \begin{figure*}[t!]
      	\centering{
      	\includegraphics[scale=0.55]{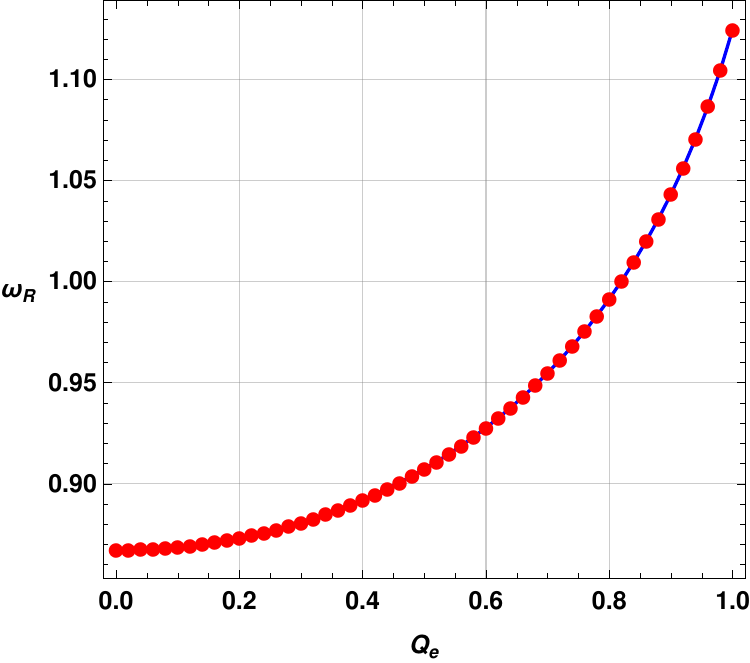}
       \includegraphics[scale=0.56]{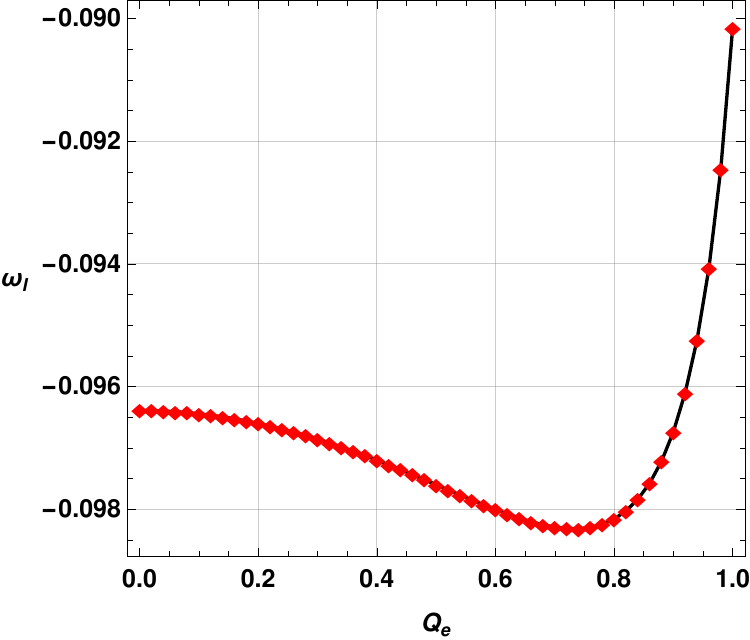}}
      	\caption{Variation of real and imaginary QNMs for static black hole using $n = 0$, $M = 1$, $l = 4$ and $A = 1$.}
      	\label{static02}
      \end{figure*}

For the static black hole case, we have used the 6th order Pad\'e averaged WKB method to calculate the QNMs using the following expression \cite{Konoplya:2019hlu, Konoplya:2011qq}:
\begin{equation}\label{qnmseqn}
\omega = \sqrt{-\, i \left[ (n + 1/2) + \sum_{k=2}^6 \bar{\Lambda}_k \right] \sqrt{-2 V_0''} + V_0},
\end{equation}

In this framework, the variable $n$ in Eq. \eqref{qnmseqn} signifies overtone numbers and can take integer values like $0, 1, 2$ etc. The value of $V_0$ is obtained by evaluating the potential function $V_s$ at the location $r_{max}$, where the potential is at its maximum. At this point, the first derivative of $V_s$ with respect to $r$ is zero, while the second derivative of $V_s$ with respect to $r$, also evaluated at $r_{max}$, is denoted as $V_0''$ \cite{Konoplya:2019hlu, Konoplya:2011qq, Konoplya:2003ii,Matyjasek:2019eeu}.

To improve the accuracy of the calculations, additional correction terms, denoted as $\bar{\Lambda}_k$, were included. These terms are explicitly defined in works \cite{Schutz:1985km,Iyer:1986np,Konoplya:2003ii,Matyjasek:2019eeu} and account for higher-order effects. These corrections are crucial for accurately predicting the oscillation frequencies of GWs in different astrophysical contexts.

The use of the Pad\'e averaging procedure, combined with these correction terms, significantly enhances the precision of the calculations. The sixth-order WKB method, with these added corrections, is an excellent tool for studying various astrophysical phenomena, including black hole mergers, neutron star oscillations, and cosmic string vibrations. However, due to the complexity of the higher-order numerical WKB method for the case of rotating black holes, we considered only the leading order corrections in the previous subsection. But, as in the case of a static black hole, the situation is comparatively simpler, one can utilise higher-order corrected WKB methods to obtain more precise values of QNMs for the black hole.

We have shown the variation of the QNMs with respect to the model parameter $Q_e$ in Fig. \ref{static02}. The charge parameter $Q_e$ non-linearly impacts the QNM spectrum. The GW frequency increases non-linearly with an increase in the value of $Q_e$. In the case of the imaginary part, we observe that the damping rate of GW increases with an increase in the value of the black hole charge parameter $Q_e$ initially but after reaching a certain peak value, it starts to decrease significantly. One may note that the variation of QNMs especially the damping or decay rate of GWs is noticeably different from the rotating case.

\section{Conclusion} \label{conc}
In this work, we have investigated the properties of black holes when the effects of the black hole charge $Q_e$ is taken into account. The latter is interpreted as a dielectric constant in vacuum arising from the one-loop of non-perturbatively quantizing the Euler-Heisenberg nonlinear electrodynamics theory. To this end, we examined its effect on the null regions, such as the horizon and ergosphere, photon-sphere, shadow radius, and observables. Interestingly, the analysis has shown considerable deviation from the Kerr and KN counterparts. The shadow cast reveals that the orbit affected by the  black hole charge $Q_e$ is the prograde orbit, suggesting that this parameter dominates near the black hole, where the gravitational field is strong.

We also investigated the QNMs for the rotating and static black hole scenarios as another means to probe the photon sphere. We found that for the rotating black hole, the variation of ring-down GWs with respect to the parameter $Q_e$ is almost similar, with a slight variation. However, in the case of damping rate, we observe a non-linear decline with an increase in the parameter $Q_e$ for the rotating black hole, and in the case of the static black hole with $a=0$, we observe an opposite scenario for smaller values of $Q_e$ and beyond a certain threshold value the damping rate decreases drastically. Interestingly, our result still confirms the correspondence between the real part of QNMs and the photon-sphere that is deeply related to the shadow radius $R_{\rm sh}$ (See. Ref.s \cite{Andersson:1995vi,Cardoso:2008bp}).

\section{Acknowledgements}
The work of G.L.  is supported by the Italian Istituto Nazionale di Fisica Nucleare (INFN) through the ``QGSKY'' project and by Ministero dell'Istruzione, Universit\`a e Ricerca (MIUR). G.L., D.J.G, R. P. and A. {\"O}. would like to acknowledge networking support of the COST Action CA21106 - COSMIC WISPers in the Dark Universe: Theory, astrophysics and experiments (CosmicWISPers), the COST Action CA22113 - Fundamental challenges in theoretical physics (THEORY-CHALLENGES), and the COST Action CA21136 - Addressing observational tensions in cosmology with systematics and fundamental physics (CosmoVerse).

\bibliography{ref}

\begin{thebibliography}{86}%
\makeatletter
\providecommand \@ifxundefined [1]{%
 \@ifx{#1\undefined}
}%
\providecommand \@ifnum [1]{%
 \ifnum #1\expandafter \@firstoftwo
 \else \expandafter \@secondoftwo
 \fi
}%
\providecommand \@ifx [1]{%
 \ifx #1\expandafter \@firstoftwo
 \else \expandafter \@secondoftwo
 \fi
}%
\providecommand \natexlab [1]{#1}%
\providecommand \enquote  [1]{``#1''}%
\providecommand \bibnamefont  [1]{#1}%
\providecommand \bibfnamefont [1]{#1}%
\providecommand \citenamefont [1]{#1}%
\providecommand \href@noop [0]{\@secondoftwo}%
\providecommand \href [0]{\begingroup \@sanitize@url \@href}%
\providecommand \@href[1]{\@@startlink{#1}\@@href}%
\providecommand \@@href[1]{\endgroup#1\@@endlink}%
\providecommand \@sanitize@url [0]{\catcode `\\12\catcode `\$12\catcode
  `\&12\catcode `\#12\catcode `\^12\catcode `\_12\catcode `\%12\relax}%
\providecommand \@@startlink[1]{}%
\providecommand \@@endlink[0]{}%
\providecommand \url  [0]{\begingroup\@sanitize@url \@url }%
\providecommand \@url [1]{\endgroup\@href {#1}{\urlprefix }}%
\providecommand \urlprefix  [0]{URL }%
\providecommand \Eprint [0]{\href }%
\providecommand \doibase [0]{http://dx.doi.org/}%
\providecommand \selectlanguage [0]{\@gobble}%
\providecommand \bibinfo  [0]{\@secondoftwo}%
\providecommand \bibfield  [0]{\@secondoftwo}%
\providecommand \translation [1]{[#1]}%
\providecommand \BibitemOpen [0]{}%
\providecommand \bibitemStop [0]{}%
\providecommand \bibitemNoStop [0]{.\EOS\space}%
\providecommand \EOS [0]{\spacefactor3000\relax}%
\providecommand \BibitemShut  [1]{\csname bibitem#1\endcsname}%
\let\auto@bib@innerbib\@empty
\bibitem [{\citenamefont {Akiyama}\ \emph
  {et~al.}(2019{\natexlab{a}})\citenamefont {Akiyama} \emph
  {et~al.}}]{EventHorizonTelescope:2019dse}%
  \BibitemOpen
  \bibfield  {author} {\bibinfo {author} {\bibfnamefont {Kazunori}\
  \bibnamefont {Akiyama}} \emph {et~al.} (\bibinfo {collaboration} {Event
  Horizon Telescope}),\ }\bibfield  {title} {\enquote {\bibinfo {title} {{First
  M87 Event Horizon Telescope Results. I. The Shadow of the Supermassive Black
  Hole}},}\ }\href {\doibase 10.3847/2041-8213/ab0ec7} {\bibfield  {journal}
  {\bibinfo  {journal} {Astrophys. J. Lett.}\ }\textbf {\bibinfo {volume}
  {875}},\ \bibinfo {pages} {L1} (\bibinfo {year} {2019}{\natexlab{a}})},\
  \Eprint {http://arxiv.org/abs/1906.11238} {arXiv:1906.11238 [astro-ph.GA]}
  \BibitemShut {NoStop}%
\bibitem [{\citenamefont {Akiyama}\ \emph
  {et~al.}(2019{\natexlab{b}})\citenamefont {Akiyama} \emph
  {et~al.}}]{EventHorizonTelescope:2019ths}%
  \BibitemOpen
  \bibfield  {author} {\bibinfo {author} {\bibfnamefont {Kazunori}\
  \bibnamefont {Akiyama}} \emph {et~al.} (\bibinfo {collaboration} {Event
  Horizon Telescope}),\ }\bibfield  {title} {\enquote {\bibinfo {title} {{First
  M87 Event Horizon Telescope Results. IV. Imaging the Central Supermassive
  Black Hole}},}\ }\href {\doibase 10.3847/2041-8213/ab0e85} {\bibfield
  {journal} {\bibinfo  {journal} {Astrophys. J. Lett.}\ }\textbf {\bibinfo
  {volume} {875}},\ \bibinfo {pages} {L4} (\bibinfo {year}
  {2019}{\natexlab{b}})},\ \Eprint {http://arxiv.org/abs/1906.11241}
  {arXiv:1906.11241 [astro-ph.GA]} \BibitemShut {NoStop}%
\bibitem [{\citenamefont {Akiyama}\ \emph
  {et~al.}(2022{\natexlab{a}})\citenamefont {Akiyama} \emph
  {et~al.}}]{EventHorizonTelescope:2022xqj}%
  \BibitemOpen
  \bibfield  {author} {\bibinfo {author} {\bibfnamefont {Kazunori}\
  \bibnamefont {Akiyama}} \emph {et~al.} (\bibinfo {collaboration} {Event
  Horizon Telescope}),\ }\bibfield  {title} {\enquote {\bibinfo {title} {{First
  Sagittarius A* Event Horizon Telescope Results. VI. Testing the Black Hole
  Metric}},}\ }\href {\doibase 10.3847/2041-8213/ac6756} {\bibfield  {journal}
  {\bibinfo  {journal} {Astrophys. J. Lett.}\ }\textbf {\bibinfo {volume}
  {930}},\ \bibinfo {pages} {L17} (\bibinfo {year} {2022}{\natexlab{a}})},\
  \Eprint {http://arxiv.org/abs/2311.09484} {arXiv:2311.09484 [astro-ph.HE]}
  \BibitemShut {NoStop}%
\bibitem [{\citenamefont {Synge}(1966)}]{Synge:1966okc}%
  \BibitemOpen
  \bibfield  {author} {\bibinfo {author} {\bibfnamefont {J.~L.}\ \bibnamefont
  {Synge}},\ }\bibfield  {title} {\enquote {\bibinfo {title} {{The Escape of
  Photons from Gravitationally Intense Stars}},}\ }\href {\doibase
  10.1093/mnras/131.3.463} {\bibfield  {journal} {\bibinfo  {journal} {Mon.
  Not. Roy. Astron. Soc.}\ }\textbf {\bibinfo {volume} {131}},\ \bibinfo
  {pages} {463--466} (\bibinfo {year} {1966})}\BibitemShut {NoStop}%
\bibitem [{\citenamefont {Luminet}(1979)}]{Luminet:1979nyg}%
  \BibitemOpen
  \bibfield  {author} {\bibinfo {author} {\bibfnamefont {J.~P.}\ \bibnamefont
  {Luminet}},\ }\bibfield  {title} {\enquote {\bibinfo {title} {{Image of a
  spherical black hole with thin accretion disk}},}\ }\href@noop {} {\bibfield
  {journal} {\bibinfo  {journal} {Astron. Astrophys.}\ }\textbf {\bibinfo
  {volume} {75}},\ \bibinfo {pages} {228--235} (\bibinfo {year}
  {1979})}\BibitemShut {NoStop}%
\bibitem [{\citenamefont {Bardeen}(1973)}]{Bardeen:1973tla}%
  \BibitemOpen
  \bibfield  {author} {\bibinfo {author} {\bibfnamefont {J.~M.}\ \bibnamefont
  {Bardeen}},\ }\bibfield  {title} {\enquote {\bibinfo {title} {{Timelike and
  null geodesics in the Kerr metric}},}\ }in\ \href@noop {} {\emph {\bibinfo
  {booktitle} {{Les Houches Summer School of Theoretical Physics}: {Black
  Holes}}}}\ (\bibinfo {year} {1973})\ pp.\ \bibinfo {pages}
  {215--240}\BibitemShut {NoStop}%
\bibitem [{\citenamefont {Akiyama}\ \emph
  {et~al.}(2022{\natexlab{b}})\citenamefont {Akiyama} \emph
  {et~al.}}]{EventHorizonTelescope:2022wkp}%
  \BibitemOpen
  \bibfield  {author} {\bibinfo {author} {\bibfnamefont {Kazunori}\
  \bibnamefont {Akiyama}} \emph {et~al.} (\bibinfo {collaboration} {Event
  Horizon Telescope}),\ }\bibfield  {title} {\enquote {\bibinfo {title} {{First
  Sagittarius A* Event Horizon Telescope Results. I. The Shadow of the
  Supermassive Black Hole in the Center of the Milky Way}},}\ }\href {\doibase
  10.3847/2041-8213/ac6674} {\bibfield  {journal} {\bibinfo  {journal}
  {Astrophys. J. Lett.}\ }\textbf {\bibinfo {volume} {930}},\ \bibinfo {pages}
  {L12} (\bibinfo {year} {2022}{\natexlab{b}})},\ \Eprint
  {http://arxiv.org/abs/2311.08680} {arXiv:2311.08680 [astro-ph.HE]}
  \BibitemShut {NoStop}%
\bibitem [{\citenamefont {Akiyama}\ \emph
  {et~al.}(2022{\natexlab{c}})\citenamefont {Akiyama} \emph
  {et~al.}}]{EventHorizonTelescope:2022wok}%
  \BibitemOpen
  \bibfield  {author} {\bibinfo {author} {\bibfnamefont {Kazunori}\
  \bibnamefont {Akiyama}} \emph {et~al.} (\bibinfo {collaboration} {Event
  Horizon Telescope}),\ }\bibfield  {title} {\enquote {\bibinfo {title} {{First
  Sagittarius A* Event Horizon Telescope Results. III. Imaging of the Galactic
  Center Supermassive Black Hole}},}\ }\href {\doibase
  10.3847/2041-8213/ac6429} {\bibfield  {journal} {\bibinfo  {journal}
  {Astrophys. J. Lett.}\ }\textbf {\bibinfo {volume} {930}},\ \bibinfo {pages}
  {L14} (\bibinfo {year} {2022}{\natexlab{c}})},\ \Eprint
  {http://arxiv.org/abs/2311.09479} {arXiv:2311.09479 [astro-ph.HE]}
  \BibitemShut {NoStop}%
\bibitem [{\citenamefont {Vagnozzi}\ \emph {et~al.}(2023)\citenamefont
  {Vagnozzi} \emph {et~al.}}]{Vagnozzi:2022moj}%
  \BibitemOpen
  \bibfield  {author} {\bibinfo {author} {\bibfnamefont {Sunny}\ \bibnamefont
  {Vagnozzi}} \emph {et~al.},\ }\bibfield  {title} {\enquote {\bibinfo {title}
  {{Horizon-scale tests of gravity theories and fundamental physics from the
  Event Horizon Telescope image of Sagittarius A$^*$}},}\ }\href {\doibase
  10.1088/1361-6382/acd97b} {\bibfield  {journal} {\bibinfo  {journal} {Class.
  Quant. Grav.}\ }\textbf {\bibinfo {volume} {40}},\ \bibinfo {pages} {165007}
  (\bibinfo {year} {2023})},\ \Eprint {http://arxiv.org/abs/2205.07787}
  {arXiv:2205.07787 [gr-qc]} \BibitemShut {NoStop}%
\bibitem [{\citenamefont {Uniyal}\ \emph {et~al.}(2023)\citenamefont {Uniyal},
  \citenamefont {Pantig},\ and\ \citenamefont {\"Ovg\"un}}]{Uniyal:2022vdu}%
  \BibitemOpen
  \bibfield  {author} {\bibinfo {author} {\bibfnamefont {Akhil}\ \bibnamefont
  {Uniyal}}, \bibinfo {author} {\bibfnamefont {Reggie~C.}\ \bibnamefont
  {Pantig}}, \ and\ \bibinfo {author} {\bibfnamefont {Ali}\ \bibnamefont
  {\"Ovg\"un}},\ }\bibfield  {title} {\enquote {\bibinfo {title} {{Probing a
  non-linear electrodynamics black hole with thin accretion disk, shadow, and
  deflection angle with M87* and Sgr A* from EHT}},}\ }\href {\doibase
  10.1016/j.dark.2023.101178} {\bibfield  {journal} {\bibinfo  {journal} {Phys.
  Dark Univ.}\ }\textbf {\bibinfo {volume} {40}},\ \bibinfo {pages} {101178}
  (\bibinfo {year} {2023})},\ \Eprint {http://arxiv.org/abs/2205.11072}
  {arXiv:2205.11072 [gr-qc]} \BibitemShut {NoStop}%
\bibitem [{\citenamefont {Zeng}\ \emph {et~al.}(2020)\citenamefont {Zeng},
  \citenamefont {Zhang},\ and\ \citenamefont {Zhang}}]{Zeng:2020dco}%
  \BibitemOpen
  \bibfield  {author} {\bibinfo {author} {\bibfnamefont {Xiao-Xiong}\
  \bibnamefont {Zeng}}, \bibinfo {author} {\bibfnamefont {Hai-Qing}\
  \bibnamefont {Zhang}}, \ and\ \bibinfo {author} {\bibfnamefont {Hongbao}\
  \bibnamefont {Zhang}},\ }\bibfield  {title} {\enquote {\bibinfo {title}
  {{Shadows and photon spheres with spherical accretions in the
  four-dimensional Gauss\textendash{}Bonnet black hole}},}\ }\href {\doibase
  10.1140/epjc/s10052-020-08449-y} {\bibfield  {journal} {\bibinfo  {journal}
  {Eur. Phys. J. C}\ }\textbf {\bibinfo {volume} {80}},\ \bibinfo {pages} {872}
  (\bibinfo {year} {2020})},\ \Eprint {http://arxiv.org/abs/2004.12074}
  {arXiv:2004.12074 [gr-qc]} \BibitemShut {NoStop}%
\bibitem [{\citenamefont {Zeng}\ \emph
  {et~al.}(2022{\natexlab{a}})\citenamefont {Zeng}, \citenamefont {Li},\ and\
  \citenamefont {He}}]{Zeng:2021dlj}%
  \BibitemOpen
  \bibfield  {author} {\bibinfo {author} {\bibfnamefont {Xiao-Xiong}\
  \bibnamefont {Zeng}}, \bibinfo {author} {\bibfnamefont {Guo-Ping}\
  \bibnamefont {Li}}, \ and\ \bibinfo {author} {\bibfnamefont {Ke-Jian}\
  \bibnamefont {He}},\ }\bibfield  {title} {\enquote {\bibinfo {title} {{The
  shadows and observational appearance of a noncommutative black hole
  surrounded by various profiles of accretions}},}\ }\href {\doibase
  10.1016/j.nuclphysb.2021.115639} {\bibfield  {journal} {\bibinfo  {journal}
  {Nucl. Phys. B}\ }\textbf {\bibinfo {volume} {974}},\ \bibinfo {pages}
  {115639} (\bibinfo {year} {2022}{\natexlab{a}})},\ \Eprint
  {http://arxiv.org/abs/2106.14478} {arXiv:2106.14478 [hep-th]} \BibitemShut
  {NoStop}%
\bibitem [{\citenamefont {Pantig}\ and\ \citenamefont
  {\"Ovg\"un}(2022{\natexlab{a}})}]{Pantig:2022whj}%
  \BibitemOpen
  \bibfield  {author} {\bibinfo {author} {\bibfnamefont {Reggie~C.}\
  \bibnamefont {Pantig}}\ and\ \bibinfo {author} {\bibfnamefont {Ali}\
  \bibnamefont {\"Ovg\"un}},\ }\bibfield  {title} {\enquote {\bibinfo {title}
  {{Dehnen halo effect on a black hole in an ultra-faint dwarf galaxy}},}\
  }\href {\doibase 10.1088/1475-7516/2022/08/056} {\bibfield  {journal}
  {\bibinfo  {journal} {JCAP}\ }\textbf {\bibinfo {volume} {08}},\ \bibinfo
  {pages} {056} (\bibinfo {year} {2022}{\natexlab{a}})},\ \Eprint
  {http://arxiv.org/abs/2202.07404} {arXiv:2202.07404 [astro-ph.GA]}
  \BibitemShut {NoStop}%
\bibitem [{\citenamefont {Pantig}\ and\ \citenamefont
  {\"Ovg\"un}(2022{\natexlab{b}})}]{Pantig:2022sjb}%
  \BibitemOpen
  \bibfield  {author} {\bibinfo {author} {\bibfnamefont {Reggie~C.}\
  \bibnamefont {Pantig}}\ and\ \bibinfo {author} {\bibfnamefont {Ali}\
  \bibnamefont {\"Ovg\"un}},\ }\bibfield  {title} {\enquote {\bibinfo {title}
  {{Black hole in quantum wave dark matter}},}\ }\href {\doibase
  10.1002/prop.202200164} {\bibfield  {journal} {\bibinfo  {journal} {Fortsch.
  Phys.}\ }\textbf {\bibinfo {volume} {2022}},\ \bibinfo {pages} {2200164}
  (\bibinfo {year} {2022}{\natexlab{b}})},\ \Eprint
  {http://arxiv.org/abs/2210.00523} {arXiv:2210.00523 [gr-qc]} \BibitemShut
  {NoStop}%
\bibitem [{\citenamefont {Zeng}\ and\ \citenamefont
  {Zhang}(2020)}]{Zeng:2020vsj}%
  \BibitemOpen
  \bibfield  {author} {\bibinfo {author} {\bibfnamefont {Xiao-Xiong}\
  \bibnamefont {Zeng}}\ and\ \bibinfo {author} {\bibfnamefont {Hai-Qing}\
  \bibnamefont {Zhang}},\ }\bibfield  {title} {\enquote {\bibinfo {title}
  {{Influence of quintessence dark energy on the shadow of black hole}},}\
  }\href {\doibase 10.1140/epjc/s10052-020-08656-7} {\bibfield  {journal}
  {\bibinfo  {journal} {Eur. Phys. J. C}\ }\textbf {\bibinfo {volume} {80}},\
  \bibinfo {pages} {1058} (\bibinfo {year} {2020})},\ \Eprint
  {http://arxiv.org/abs/2007.06333} {arXiv:2007.06333 [gr-qc]} \BibitemShut
  {NoStop}%
\bibitem [{\citenamefont {\"Ovg\"un}\ and\ \citenamefont
  {Sakall\i{}}(2020)}]{Ovgun:2020gjz}%
  \BibitemOpen
  \bibfield  {author} {\bibinfo {author} {\bibfnamefont {Ali}\ \bibnamefont
  {\"Ovg\"un}}\ and\ \bibinfo {author} {\bibfnamefont {\.Izzet}\ \bibnamefont
  {Sakall\i{}}},\ }\bibfield  {title} {\enquote {\bibinfo {title} {{Testing
  generalized Einstein\textendash{}Cartan\textendash{}Kibble\textendash{}Sciama
  gravity using weak deflection angle and shadow cast}},}\ }\href {\doibase
  10.1088/1361-6382/abb579} {\bibfield  {journal} {\bibinfo  {journal} {Class.
  Quant. Grav.}\ }\textbf {\bibinfo {volume} {37}},\ \bibinfo {pages} {225003}
  (\bibinfo {year} {2020})},\ \Eprint {http://arxiv.org/abs/2005.00982}
  {arXiv:2005.00982 [gr-qc]} \BibitemShut {NoStop}%
\bibitem [{\citenamefont {Kuang}\ and\ \citenamefont
  {\"Ovg\"un}(2022)}]{Kuang:2022xjp}%
  \BibitemOpen
  \bibfield  {author} {\bibinfo {author} {\bibfnamefont {Xiao-Mei}\
  \bibnamefont {Kuang}}\ and\ \bibinfo {author} {\bibfnamefont {Ali}\
  \bibnamefont {\"Ovg\"un}},\ }\bibfield  {title} {\enquote {\bibinfo {title}
  {{Strong gravitational lensing and shadow constraint from M87* of slowly
  rotating Kerr-like black hole}},}\ }\href {\doibase
  10.1016/j.aop.2022.169147} {\bibfield  {journal} {\bibinfo  {journal} {Annals
  Phys.}\ }\textbf {\bibinfo {volume} {447}},\ \bibinfo {pages} {169147}
  (\bibinfo {year} {2022})},\ \Eprint {http://arxiv.org/abs/2205.11003}
  {arXiv:2205.11003 [gr-qc]} \BibitemShut {NoStop}%
\bibitem [{\citenamefont {Mustafa}\ \emph {et~al.}(2022)\citenamefont
  {Mustafa}, \citenamefont {Atamurotov}, \citenamefont {Hussain}, \citenamefont
  {Shaymatov},\ and\ \citenamefont {\"Ovg\"un}}]{Mustafa:2022xod}%
  \BibitemOpen
  \bibfield  {author} {\bibinfo {author} {\bibfnamefont {Ghulam}\ \bibnamefont
  {Mustafa}}, \bibinfo {author} {\bibfnamefont {Farruh}\ \bibnamefont
  {Atamurotov}}, \bibinfo {author} {\bibfnamefont {Ibrar}\ \bibnamefont
  {Hussain}}, \bibinfo {author} {\bibfnamefont {Sanjar}\ \bibnamefont
  {Shaymatov}}, \ and\ \bibinfo {author} {\bibfnamefont {Ali}\ \bibnamefont
  {\"Ovg\"un}},\ }\bibfield  {title} {\enquote {\bibinfo {title} {{Shadows and
  gravitational weak lensing by the Schwarzschild black hole in the string
  cloud background with quintessential field*}},}\ }\href {\doibase
  10.1088/1674-1137/ac917f} {\bibfield  {journal} {\bibinfo  {journal} {Chin.
  Phys. C}\ }\textbf {\bibinfo {volume} {46}},\ \bibinfo {pages} {125107}
  (\bibinfo {year} {2022})},\ \Eprint {http://arxiv.org/abs/2207.07608}
  {arXiv:2207.07608 [gr-qc]} \BibitemShut {NoStop}%
\bibitem [{\citenamefont {Kumaran}\ and\ \citenamefont
  {\"Ovg\"un}(2022)}]{Kumaran:2022soh}%
  \BibitemOpen
  \bibfield  {author} {\bibinfo {author} {\bibfnamefont {Yashmitha}\
  \bibnamefont {Kumaran}}\ and\ \bibinfo {author} {\bibfnamefont {Ali}\
  \bibnamefont {\"Ovg\"un}},\ }\bibfield  {title} {\enquote {\bibinfo {title}
  {{Deflection Angle and Shadow of the Reissner\textendash{}Nordstr\"om Black
  Hole with Higher-Order Magnetic Correction in Einstein-Nonlinear-Maxwell
  Fields}},}\ }\href {\doibase 10.3390/sym14102054} {\bibfield  {journal}
  {\bibinfo  {journal} {Symmetry}\ }\textbf {\bibinfo {volume} {14}},\ \bibinfo
  {pages} {2054} (\bibinfo {year} {2022})},\ \Eprint
  {http://arxiv.org/abs/2210.00468} {arXiv:2210.00468 [gr-qc]} \BibitemShut
  {NoStop}%
\bibitem [{\citenamefont {Cimdiker}\ \emph {et~al.}(2021)\citenamefont
  {Cimdiker}, \citenamefont {Demir},\ and\ \citenamefont
  {\"Ovg\"un}}]{Cimdiker:2021cpz}%
  \BibitemOpen
  \bibfield  {author} {\bibinfo {author} {\bibfnamefont {\.Irfan}\ \bibnamefont
  {Cimdiker}}, \bibinfo {author} {\bibfnamefont {Durmus}\ \bibnamefont
  {Demir}}, \ and\ \bibinfo {author} {\bibfnamefont {Ali}\ \bibnamefont
  {\"Ovg\"un}},\ }\bibfield  {title} {\enquote {\bibinfo {title} {{Black hole
  shadow in symmergent gravity}},}\ }\href {\doibase
  10.1016/j.dark.2021.100900} {\bibfield  {journal} {\bibinfo  {journal} {Phys.
  Dark Univ.}\ }\textbf {\bibinfo {volume} {34}},\ \bibinfo {pages} {100900}
  (\bibinfo {year} {2021})},\ \Eprint {http://arxiv.org/abs/2110.11904}
  {arXiv:2110.11904 [gr-qc]} \BibitemShut {NoStop}%
\bibitem [{\citenamefont {Okyay}\ and\ \citenamefont
  {\"Ovg\"un}(2022)}]{Okyay:2021nnh}%
  \BibitemOpen
  \bibfield  {author} {\bibinfo {author} {\bibfnamefont {Mert}\ \bibnamefont
  {Okyay}}\ and\ \bibinfo {author} {\bibfnamefont {Ali}\ \bibnamefont
  {\"Ovg\"un}},\ }\bibfield  {title} {\enquote {\bibinfo {title} {{Nonlinear
  electrodynamics effects on the black hole shadow, deflection angle,
  quasinormal modes and greybody factors}},}\ }\href {\doibase
  10.1088/1475-7516/2022/01/009} {\bibfield  {journal} {\bibinfo  {journal}
  {JCAP}\ }\textbf {\bibinfo {volume} {01}},\ \bibinfo {pages} {009} (\bibinfo
  {year} {2022})},\ \Eprint {http://arxiv.org/abs/2108.07766} {arXiv:2108.07766
  [gr-qc]} \BibitemShut {NoStop}%
\bibitem [{\citenamefont {Atamurotov}\ \emph {et~al.}(2023)\citenamefont
  {Atamurotov}, \citenamefont {Hussain}, \citenamefont {Mustafa},\ and\
  \citenamefont {\"Ovg\"un}}]{Atamurotov:2022knb}%
  \BibitemOpen
  \bibfield  {author} {\bibinfo {author} {\bibfnamefont {Farruh}\ \bibnamefont
  {Atamurotov}}, \bibinfo {author} {\bibfnamefont {Ibrar}\ \bibnamefont
  {Hussain}}, \bibinfo {author} {\bibfnamefont {Ghulam}\ \bibnamefont
  {Mustafa}}, \ and\ \bibinfo {author} {\bibfnamefont {Ali}\ \bibnamefont
  {\"Ovg\"un}},\ }\bibfield  {title} {\enquote {\bibinfo {title} {{Weak
  deflection angle and shadow cast by the charged-Kiselev black hole with cloud
  of strings in plasma*}},}\ }\href {\doibase 10.1088/1674-1137/ac9fbb}
  {\bibfield  {journal} {\bibinfo  {journal} {Chin. Phys. C}\ }\textbf
  {\bibinfo {volume} {47}},\ \bibinfo {pages} {025102} (\bibinfo {year}
  {2023})}\BibitemShut {NoStop}%
\bibitem [{\citenamefont {Olmo}\ \emph {et~al.}(2023)\citenamefont {Olmo},
  \citenamefont {Rosa}, \citenamefont {Rubiera-Garcia},\ and\ \citenamefont
  {Saez-Chillon~Gomez}}]{Olmo:2023lil}%
  \BibitemOpen
  \bibfield  {author} {\bibinfo {author} {\bibfnamefont {Gonzalo~J.}\
  \bibnamefont {Olmo}}, \bibinfo {author} {\bibfnamefont {Joao~Luis}\
  \bibnamefont {Rosa}}, \bibinfo {author} {\bibfnamefont {Diego}\ \bibnamefont
  {Rubiera-Garcia}}, \ and\ \bibinfo {author} {\bibfnamefont {Diego}\
  \bibnamefont {Saez-Chillon~Gomez}},\ }\bibfield  {title} {\enquote {\bibinfo
  {title} {{Shadows and photon rings of regular black holes and geonic
  horizonless compact objects}},}\ }\href {\doibase 10.1088/1361-6382/aceacd}
  {\bibfield  {journal} {\bibinfo  {journal} {Class. Quant. Grav.}\ }\textbf
  {\bibinfo {volume} {40}},\ \bibinfo {pages} {174002} (\bibinfo {year}
  {2023})},\ \Eprint {http://arxiv.org/abs/2302.12064} {arXiv:2302.12064
  [gr-qc]} \BibitemShut {NoStop}%
\bibitem [{\citenamefont {Asuk\"ula}\ \emph {et~al.}(2024)\citenamefont
  {Asuk\"ula}, \citenamefont {Hohmann}, \citenamefont {Karanasou},
  \citenamefont {Bahamonde}, \citenamefont {Pfeifer},\ and\ \citenamefont
  {Rosa}}]{Asukula:2023akj}%
  \BibitemOpen
  \bibfield  {author} {\bibinfo {author} {\bibfnamefont {Helen}\ \bibnamefont
  {Asuk\"ula}}, \bibinfo {author} {\bibfnamefont {Manuel}\ \bibnamefont
  {Hohmann}}, \bibinfo {author} {\bibfnamefont {Vasiliki}\ \bibnamefont
  {Karanasou}}, \bibinfo {author} {\bibfnamefont {Sebastian}\ \bibnamefont
  {Bahamonde}}, \bibinfo {author} {\bibfnamefont {Christian}\ \bibnamefont
  {Pfeifer}}, \ and\ \bibinfo {author} {\bibfnamefont {Jo\~ao~Lu\'\i{}s}\
  \bibnamefont {Rosa}},\ }\bibfield  {title} {\enquote {\bibinfo {title}
  {{Spherically symmetric vacuum solutions in one-parameter new general
  relativity and their phenomenology}},}\ }\href {\doibase
  10.1103/PhysRevD.109.064027} {\bibfield  {journal} {\bibinfo  {journal}
  {Phys. Rev. D}\ }\textbf {\bibinfo {volume} {109}},\ \bibinfo {pages}
  {064027} (\bibinfo {year} {2024})},\ \Eprint
  {http://arxiv.org/abs/2311.17999} {arXiv:2311.17999 [gr-qc]} \BibitemShut
  {NoStop}%
\bibitem [{\citenamefont {Zeng}\ \emph {et~al.}(2023)\citenamefont {Zeng},
  \citenamefont {He}, \citenamefont {Pu}, \citenamefont {Li},\ and\
  \citenamefont {Jiang}}]{Zeng:2023zlf}%
  \BibitemOpen
  \bibfield  {author} {\bibinfo {author} {\bibfnamefont {Xiao-Xiong}\
  \bibnamefont {Zeng}}, \bibinfo {author} {\bibfnamefont {Ke-Jian}\
  \bibnamefont {He}}, \bibinfo {author} {\bibfnamefont {Jin}\ \bibnamefont
  {Pu}}, \bibinfo {author} {\bibfnamefont {Guo-ping}\ \bibnamefont {Li}}, \
  and\ \bibinfo {author} {\bibfnamefont {Qing-Quan}\ \bibnamefont {Jiang}},\
  }\bibfield  {title} {\enquote {\bibinfo {title} {{Holographic Einstein rings
  of a Gauss\textendash{}Bonnet AdS black hole}},}\ }\href {\doibase
  10.1140/epjc/s10052-023-12079-5} {\bibfield  {journal} {\bibinfo  {journal}
  {Eur. Phys. J. C}\ }\textbf {\bibinfo {volume} {83}},\ \bibinfo {pages} {897}
  (\bibinfo {year} {2023})},\ \Eprint {http://arxiv.org/abs/2302.03692}
  {arXiv:2302.03692 [gr-qc]} \BibitemShut {NoStop}%
\bibitem [{\citenamefont {Vagnozzi}\ and\ \citenamefont
  {Visinelli}(2019)}]{Vagnozzi:2019apd}%
  \BibitemOpen
  \bibfield  {author} {\bibinfo {author} {\bibfnamefont {Sunny}\ \bibnamefont
  {Vagnozzi}}\ and\ \bibinfo {author} {\bibfnamefont {Luca}\ \bibnamefont
  {Visinelli}},\ }\bibfield  {title} {\enquote {\bibinfo {title} {{Hunting for
  extra dimensions in the shadow of M87*}},}\ }\href {\doibase
  10.1103/PhysRevD.100.024020} {\bibfield  {journal} {\bibinfo  {journal}
  {Phys. Rev. D}\ }\textbf {\bibinfo {volume} {100}},\ \bibinfo {pages}
  {024020} (\bibinfo {year} {2019})},\ \Eprint
  {http://arxiv.org/abs/1905.12421} {arXiv:1905.12421 [gr-qc]} \BibitemShut
  {NoStop}%
\bibitem [{\citenamefont {Yajima}\ and\ \citenamefont
  {Tamaki}(2001)}]{Yajima:2000kw}%
  \BibitemOpen
  \bibfield  {author} {\bibinfo {author} {\bibfnamefont {Hiroki}\ \bibnamefont
  {Yajima}}\ and\ \bibinfo {author} {\bibfnamefont {Takashi}\ \bibnamefont
  {Tamaki}},\ }\bibfield  {title} {\enquote {\bibinfo {title} {{Black hole
  solutions in Euler-Heisenberg theory}},}\ }\href {\doibase
  10.1103/PhysRevD.63.064007} {\bibfield  {journal} {\bibinfo  {journal} {Phys.
  Rev. D}\ }\textbf {\bibinfo {volume} {63}},\ \bibinfo {pages} {064007}
  (\bibinfo {year} {2001})},\ \Eprint {http://arxiv.org/abs/gr-qc/0005016}
  {arXiv:gr-qc/0005016} \BibitemShut {NoStop}%
\bibitem [{\citenamefont {Ruffini}\ \emph {et~al.}(2013)\citenamefont
  {Ruffini}, \citenamefont {Wu},\ and\ \citenamefont {Xue}}]{Ruffini:2013hia}%
  \BibitemOpen
  \bibfield  {author} {\bibinfo {author} {\bibfnamefont {Remo}\ \bibnamefont
  {Ruffini}}, \bibinfo {author} {\bibfnamefont {Yuan-Bin}\ \bibnamefont {Wu}},
  \ and\ \bibinfo {author} {\bibfnamefont {She-Sheng}\ \bibnamefont {Xue}},\
  }\bibfield  {title} {\enquote {\bibinfo {title} {{Einstein-Euler-Heisenberg
  Theory and charged black holes}},}\ }\href {\doibase
  10.1103/PhysRevD.88.085004} {\bibfield  {journal} {\bibinfo  {journal} {Phys.
  Rev. D}\ }\textbf {\bibinfo {volume} {88}},\ \bibinfo {pages} {085004}
  (\bibinfo {year} {2013})},\ \Eprint {http://arxiv.org/abs/1307.4951}
  {arXiv:1307.4951 [hep-th]} \BibitemShut {NoStop}%
\bibitem [{\citenamefont {Bret\'on}\ \emph {et~al.}(2019)\citenamefont
  {Bret\'on}, \citenamefont {L\"ammerzahl},\ and\ \citenamefont
  {Mac\'\i{}as}}]{Breton:2019arv}%
  \BibitemOpen
  \bibfield  {author} {\bibinfo {author} {\bibfnamefont {Nora}\ \bibnamefont
  {Bret\'on}}, \bibinfo {author} {\bibfnamefont {Claus}\ \bibnamefont
  {L\"ammerzahl}}, \ and\ \bibinfo {author} {\bibfnamefont {Alfredo}\
  \bibnamefont {Mac\'\i{}as}},\ }\bibfield  {title} {\enquote {\bibinfo {title}
  {{Rotating black holes in the
  Einstein\textendash{}Euler\textendash{}Heisenberg theory}},}\ }\href
  {\doibase 10.1088/1361-6382/ab5169} {\bibfield  {journal} {\bibinfo
  {journal} {Class. Quant. Grav.}\ }\textbf {\bibinfo {volume} {36}},\ \bibinfo
  {pages} {235022} (\bibinfo {year} {2019})}\BibitemShut {NoStop}%
\bibitem [{\citenamefont {Bret\'on}\ \emph {et~al.}(2022)\citenamefont
  {Bret\'on}, \citenamefont {L\"ammerzahl},\ and\ \citenamefont
  {Mac\'\i{}as}}]{Breton:2022fch}%
  \BibitemOpen
  \bibfield  {author} {\bibinfo {author} {\bibfnamefont {Nora}\ \bibnamefont
  {Bret\'on}}, \bibinfo {author} {\bibfnamefont {Claus}\ \bibnamefont
  {L\"ammerzahl}}, \ and\ \bibinfo {author} {\bibfnamefont {Alfredo}\
  \bibnamefont {Mac\'\i{}as}},\ }\bibfield  {title} {\enquote {\bibinfo {title}
  {{Rotating structure of the Euler-Heisenberg black hole}},}\ }\href {\doibase
  10.1103/PhysRevD.105.104046} {\bibfield  {journal} {\bibinfo  {journal}
  {Phys. Rev. D}\ }\textbf {\bibinfo {volume} {105}},\ \bibinfo {pages}
  {104046} (\bibinfo {year} {2022})}\BibitemShut {NoStop}%
\bibitem [{\citenamefont {Zhong}\ \emph {et~al.}(2021)\citenamefont {Zhong},
  \citenamefont {Hu}, \citenamefont {Yan}, \citenamefont {Guo},\ and\
  \citenamefont {Chen}}]{Zhong:2021mty}%
  \BibitemOpen
  \bibfield  {author} {\bibinfo {author} {\bibfnamefont {Zhen}\ \bibnamefont
  {Zhong}}, \bibinfo {author} {\bibfnamefont {Zezhou}\ \bibnamefont {Hu}},
  \bibinfo {author} {\bibfnamefont {Haopeng}\ \bibnamefont {Yan}}, \bibinfo
  {author} {\bibfnamefont {Minyong}\ \bibnamefont {Guo}}, \ and\ \bibinfo
  {author} {\bibfnamefont {Bin}\ \bibnamefont {Chen}},\ }\bibfield  {title}
  {\enquote {\bibinfo {title} {{QED effects on Kerr black hole shadows immersed
  in uniform magnetic fields}},}\ }\href {\doibase 10.1103/PhysRevD.104.104028}
  {\bibfield  {journal} {\bibinfo  {journal} {Phys. Rev. D}\ }\textbf {\bibinfo
  {volume} {104}},\ \bibinfo {pages} {104028} (\bibinfo {year} {2021})},\
  \Eprint {http://arxiv.org/abs/2108.06140} {arXiv:2108.06140 [gr-qc]}
  \BibitemShut {NoStop}%
\bibitem [{\citenamefont {Amaro}\ \emph {et~al.}(2023)\citenamefont {Amaro},
  \citenamefont {L\"ammerzahl},\ and\ \citenamefont
  {Mac\'\i{}as}}]{Amaro:2023ull}%
  \BibitemOpen
  \bibfield  {author} {\bibinfo {author} {\bibfnamefont {Daniel}\ \bibnamefont
  {Amaro}}, \bibinfo {author} {\bibfnamefont {Claus}\ \bibnamefont
  {L\"ammerzahl}}, \ and\ \bibinfo {author} {\bibfnamefont {Alfredo}\
  \bibnamefont {Mac\'\i{}as}},\ }\bibfield  {title} {\enquote {\bibinfo {title}
  {{Particle motion in the Einstein-Euler-Heisenberg rotating black hole
  spacetime}},}\ }\href {\doibase 10.1103/PhysRevD.107.084040} {\bibfield
  {journal} {\bibinfo  {journal} {Phys. Rev. D}\ }\textbf {\bibinfo {volume}
  {107}},\ \bibinfo {pages} {084040} (\bibinfo {year} {2023})}\BibitemShut
  {NoStop}%
\bibitem [{\citenamefont {Zeng}\ \emph
  {et~al.}(2022{\natexlab{b}})\citenamefont {Zeng}, \citenamefont {He},
  \citenamefont {Li}, \citenamefont {Liang},\ and\ \citenamefont
  {Guo}}]{Zeng:2022pvb}%
  \BibitemOpen
  \bibfield  {author} {\bibinfo {author} {\bibfnamefont {Xiao-Xiong}\
  \bibnamefont {Zeng}}, \bibinfo {author} {\bibfnamefont {Ke-Jian}\
  \bibnamefont {He}}, \bibinfo {author} {\bibfnamefont {Guo-Ping}\ \bibnamefont
  {Li}}, \bibinfo {author} {\bibfnamefont {En-Wei}\ \bibnamefont {Liang}}, \
  and\ \bibinfo {author} {\bibfnamefont {Sen}\ \bibnamefont {Guo}},\ }\bibfield
   {title} {\enquote {\bibinfo {title} {{QED and accretion flow models effect
  on optical appearance of Euler\textendash{}Heisenberg black holes}},}\ }\href
  {\doibase 10.1140/epjc/s10052-022-10733-y} {\bibfield  {journal} {\bibinfo
  {journal} {Eur. Phys. J. C}\ }\textbf {\bibinfo {volume} {82}},\ \bibinfo
  {pages} {764} (\bibinfo {year} {2022}{\natexlab{b}})},\ \Eprint
  {http://arxiv.org/abs/2209.05938} {arXiv:2209.05938 [gr-qc]} \BibitemShut
  {NoStop}%
\bibitem [{\citenamefont {Bret\'on}\ and\ \citenamefont
  {L\'opez}(2021)}]{Breton:2021mju}%
  \BibitemOpen
  \bibfield  {author} {\bibinfo {author} {\bibfnamefont {Nora}\ \bibnamefont
  {Bret\'on}}\ and\ \bibinfo {author} {\bibfnamefont {L.~A.}\ \bibnamefont
  {L\'opez}},\ }\bibfield  {title} {\enquote {\bibinfo {title} {{Birefringence
  and quasinormal modes of the Einstein-Euler-Heisenberg black hole}},}\ }\href
  {\doibase 10.1103/PhysRevD.104.024064} {\bibfield  {journal} {\bibinfo
  {journal} {Phys. Rev. D}\ }\textbf {\bibinfo {volume} {104}},\ \bibinfo
  {pages} {024064} (\bibinfo {year} {2021})},\ \Eprint
  {http://arxiv.org/abs/2105.12283} {arXiv:2105.12283 [gr-qc]} \BibitemShut
  {NoStop}%
\bibitem [{\citenamefont {Luo}\ and\ \citenamefont {Li}(2022)}]{Luo:2022gdz}%
  \BibitemOpen
  \bibfield  {author} {\bibinfo {author} {\bibfnamefont {Zhi}\ \bibnamefont
  {Luo}}\ and\ \bibinfo {author} {\bibfnamefont {Jin}\ \bibnamefont {Li}},\
  }\bibfield  {title} {\enquote {\bibinfo {title} {{Gravitational perturbations
  of the Einstein-Euler-Heisenberg black hole *}},}\ }\href {\doibase
  10.1088/1674-1137/ac6574} {\bibfield  {journal} {\bibinfo  {journal} {Chin.
  Phys. C}\ }\textbf {\bibinfo {volume} {46}},\ \bibinfo {pages} {085107}
  (\bibinfo {year} {2022})}\BibitemShut {NoStop}%
\bibitem [{\citenamefont {Dai}\ \emph {et~al.}(2023)\citenamefont {Dai},
  \citenamefont {Zhao},\ and\ \citenamefont {Zhang}}]{Dai:2022mko}%
  \BibitemOpen
  \bibfield  {author} {\bibinfo {author} {\bibfnamefont {Heng}\ \bibnamefont
  {Dai}}, \bibinfo {author} {\bibfnamefont {Zixu}\ \bibnamefont {Zhao}}, \ and\
  \bibinfo {author} {\bibfnamefont {Shuhang}\ \bibnamefont {Zhang}},\
  }\bibfield  {title} {\enquote {\bibinfo {title} {{Thermodynamic phase
  transition of Euler-Heisenberg-AdS black hole on free energy landscape}},}\
  }\href {\doibase 10.1016/j.nuclphysb.2023.116219} {\bibfield  {journal}
  {\bibinfo  {journal} {Nucl. Phys. B}\ }\textbf {\bibinfo {volume} {991}},\
  \bibinfo {pages} {116219} (\bibinfo {year} {2023})},\ \Eprint
  {http://arxiv.org/abs/2202.14007} {arXiv:2202.14007 [gr-qc]} \BibitemShut
  {NoStop}%
\bibitem [{\citenamefont {Feng}\ and\ \citenamefont
  {Nie}(2022)}]{Feng:2022otu}%
  \BibitemOpen
  \bibfield  {author} {\bibinfo {author} {\bibfnamefont {Yuanyuan}\
  \bibnamefont {Feng}}\ and\ \bibinfo {author} {\bibfnamefont {Weifu}\
  \bibnamefont {Nie}},\ }\bibfield  {title} {\enquote {\bibinfo {title} {{The
  Correspondence Between Shadow and the Test Field in a
  Einstein-Euler-Heisenberg Black Hole}},}\ }\href {\doibase
  10.1007/s10773-022-05205-8} {\bibfield  {journal} {\bibinfo  {journal} {Int.
  J. Theor. Phys.}\ }\textbf {\bibinfo {volume} {61}},\ \bibinfo {pages} {223}
  (\bibinfo {year} {2022})}\BibitemShut {NoStop}%
\bibitem [{\citenamefont {Maceda}\ and\ \citenamefont
  {Mac\'\i{}as}(2019)}]{Maceda:2018zim}%
  \BibitemOpen
  \bibfield  {author} {\bibinfo {author} {\bibfnamefont {Marco}\ \bibnamefont
  {Maceda}}\ and\ \bibinfo {author} {\bibfnamefont {Alfredo}\ \bibnamefont
  {Mac\'\i{}as}},\ }\bibfield  {title} {\enquote {\bibinfo {title}
  {{Non-commutative inspired black holes in Euler\textendash{}Heisenberg
  non-linear electrodynamics}},}\ }\href {\doibase
  10.1016/j.physletb.2018.11.048} {\bibfield  {journal} {\bibinfo  {journal}
  {Phys. Lett. B}\ }\textbf {\bibinfo {volume} {788}},\ \bibinfo {pages}
  {446--452} (\bibinfo {year} {2019})},\ \Eprint
  {http://arxiv.org/abs/1807.05269} {arXiv:1807.05269 [gr-qc]} \BibitemShut
  {NoStop}%
\bibitem [{\citenamefont {Maceda}\ \emph {et~al.}(2021)\citenamefont {Maceda},
  \citenamefont {Macias},\ and\ \citenamefont
  {Martinez-Carbajal}}]{Maceda:2020rpv}%
  \BibitemOpen
  \bibfield  {author} {\bibinfo {author} {\bibfnamefont {Marco}\ \bibnamefont
  {Maceda}}, \bibinfo {author} {\bibfnamefont {Alfredo}\ \bibnamefont
  {Macias}}, \ and\ \bibinfo {author} {\bibfnamefont {Daniel}\ \bibnamefont
  {Martinez-Carbajal}},\ }\bibfield  {title} {\enquote {\bibinfo {title}
  {{Shadow of a noncommutative-inspired
  Einstein\textendash{}Euler\textendash{}Heisenberg black hole}},}\ }\href
  {\doibase 10.1142/S0217751X21501918} {\bibfield  {journal} {\bibinfo
  {journal} {Int. J. Mod. Phys. A}\ }\textbf {\bibinfo {volume} {36}},\
  \bibinfo {pages} {2150191} (\bibinfo {year} {2021})},\ \Eprint
  {http://arxiv.org/abs/2008.07040} {arXiv:2008.07040 [gr-qc]} \BibitemShut
  {NoStop}%
\bibitem [{\citenamefont {Rehman}\ \emph {et~al.}(2023)\citenamefont {Rehman},
  \citenamefont {Abbas}, \citenamefont {Zhu},\ and\ \citenamefont
  {Mustafa}}]{Rehman:2023hro}%
  \BibitemOpen
  \bibfield  {author} {\bibinfo {author} {\bibfnamefont {H.}~\bibnamefont
  {Rehman}}, \bibinfo {author} {\bibfnamefont {G.}~\bibnamefont {Abbas}},
  \bibinfo {author} {\bibfnamefont {Tao}\ \bibnamefont {Zhu}}, \ and\ \bibinfo
  {author} {\bibfnamefont {G.}~\bibnamefont {Mustafa}},\ }\bibfield  {title}
  {\enquote {\bibinfo {title} {{Matter accretion onto the magnetically charged
  Euler\textendash{}Heisenberg black hole with scalar hair}},}\ }\href
  {\doibase 10.1140/epjc/s10052-023-12033-5} {\bibfield  {journal} {\bibinfo
  {journal} {Eur. Phys. J. C}\ }\textbf {\bibinfo {volume} {83}},\ \bibinfo
  {pages} {856} (\bibinfo {year} {2023})},\ \Eprint
  {http://arxiv.org/abs/2307.16155} {arXiv:2307.16155 [gr-qc]} \BibitemShut
  {NoStop}%
\bibitem [{\citenamefont {Mushtaq}\ \emph {et~al.}(2024)\citenamefont
  {Mushtaq}, \citenamefont {Tiecheng}, \citenamefont {Ditta}, \citenamefont
  {Atamurotov}, \citenamefont {Abduvokhidov},\ and\ \citenamefont
  {Asalkhon}}]{Mushtaq:2024cap}%
  \BibitemOpen
  \bibfield  {author} {\bibinfo {author} {\bibfnamefont {Farzan}\ \bibnamefont
  {Mushtaq}}, \bibinfo {author} {\bibfnamefont {Xia}\ \bibnamefont {Tiecheng}},
  \bibinfo {author} {\bibfnamefont {Allah}\ \bibnamefont {Ditta}}, \bibinfo
  {author} {\bibfnamefont {Farruh}\ \bibnamefont {Atamurotov}}, \bibinfo
  {author} {\bibfnamefont {Alisher}\ \bibnamefont {Abduvokhidov}}, \ and\
  \bibinfo {author} {\bibfnamefont {Alimova}\ \bibnamefont {Asalkhon}},\
  }\bibfield  {title} {\enquote {\bibinfo {title} {{Weak gravitational lensing
  and fundamental frequencies of
  Einstein\textendash{}Euler\textendash{}Heisenberg black hole}},}\ }\href
  {\doibase 10.1016/j.newast.2023.102185} {\bibfield  {journal} {\bibinfo
  {journal} {New Astron.}\ }\textbf {\bibinfo {volume} {108}},\ \bibinfo
  {pages} {102185} (\bibinfo {year} {2024})}\BibitemShut {NoStop}%
\bibitem [{\citenamefont {Hu}\ \emph {et~al.}(2021)\citenamefont {Hu},
  \citenamefont {Zhong}, \citenamefont {Li}, \citenamefont {Guo},\ and\
  \citenamefont {Chen}}]{Hu:2020usx}%
  \BibitemOpen
  \bibfield  {author} {\bibinfo {author} {\bibfnamefont {Zezhou}\ \bibnamefont
  {Hu}}, \bibinfo {author} {\bibfnamefont {Zhen}\ \bibnamefont {Zhong}},
  \bibinfo {author} {\bibfnamefont {Peng-Cheng}\ \bibnamefont {Li}}, \bibinfo
  {author} {\bibfnamefont {Minyong}\ \bibnamefont {Guo}}, \ and\ \bibinfo
  {author} {\bibfnamefont {Bin}\ \bibnamefont {Chen}},\ }\bibfield  {title}
  {\enquote {\bibinfo {title} {{QED effect on a black hole shadow}},}\ }\href
  {\doibase 10.1103/PhysRevD.103.044057} {\bibfield  {journal} {\bibinfo
  {journal} {Phys. Rev. D}\ }\textbf {\bibinfo {volume} {103}},\ \bibinfo
  {pages} {044057} (\bibinfo {year} {2021})},\ \Eprint
  {http://arxiv.org/abs/2012.07022} {arXiv:2012.07022 [gr-qc]} \BibitemShut
  {NoStop}%
\bibitem [{\citenamefont {Abbott~et al.}(2016)}]{PhysRevLett.116.061102}%
  \BibitemOpen
  \bibfield  {author} {\bibinfo {author} {\bibfnamefont {B.~P.}\ \bibnamefont
  {Abbott~et al.}} (\bibinfo {collaboration} {LIGO Scientific Collaboration and
  Virgo Collaboration}),\ }\bibfield  {title} {\enquote {\bibinfo {title}
  {Observation of gravitational waves from a binary black hole merger},}\
  }\href {\doibase 10.1103/PhysRevLett.116.061102} {\bibfield  {journal}
  {\bibinfo  {journal} {Phys. Rev. Lett.}\ }\textbf {\bibinfo {volume} {116}},\
  \bibinfo {pages} {061102} (\bibinfo {year} {2016})}\BibitemShut {NoStop}%
\bibitem [{\citenamefont {Vishveshwara}(1970)}]{Vishveshwara:1970zz}%
  \BibitemOpen
  \bibfield  {author} {\bibinfo {author} {\bibfnamefont {C.~V.}\ \bibnamefont
  {Vishveshwara}},\ }\bibfield  {title} {\enquote {\bibinfo {title}
  {{Scattering of Gravitational Radiation by a Schwarzschild Black-hole}},}\
  }\href {\doibase 10.1038/227936a0} {\bibfield  {journal} {\bibinfo  {journal}
  {Nature}\ }\textbf {\bibinfo {volume} {227}},\ \bibinfo {pages} {936--938}
  (\bibinfo {year} {1970})}\BibitemShut {NoStop}%
\bibitem [{\citenamefont {Press}(1971)}]{Press:1971wr}%
  \BibitemOpen
  \bibfield  {author} {\bibinfo {author} {\bibfnamefont {William~H.}\
  \bibnamefont {Press}},\ }\bibfield  {title} {\enquote {\bibinfo {title}
  {{Long Wave Trains of Gravitational Waves from a Vibrating Black Hole}},}\
  }\href {\doibase 10.1086/180849} {\bibfield  {journal} {\bibinfo  {journal}
  {Astrophys. J. Lett.}\ }\textbf {\bibinfo {volume} {170}},\ \bibinfo {pages}
  {L105--L108} (\bibinfo {year} {1971})}\BibitemShut {NoStop}%
\bibitem [{\citenamefont {Kokkotas}\ and\ \citenamefont
  {Schmidt}(1999)}]{Kokkotas:1999bd}%
  \BibitemOpen
  \bibfield  {author} {\bibinfo {author} {\bibfnamefont {Kostas~D.}\
  \bibnamefont {Kokkotas}}\ and\ \bibinfo {author} {\bibfnamefont {Bernd~G.}\
  \bibnamefont {Schmidt}},\ }\bibfield  {title} {\enquote {\bibinfo {title}
  {{Quasinormal modes of stars and black holes}},}\ }\href {\doibase
  10.12942/lrr-1999-2} {\bibfield  {journal} {\bibinfo  {journal} {Living Rev.
  Rel.}\ }\textbf {\bibinfo {volume} {2}},\ \bibinfo {pages} {2} (\bibinfo
  {year} {1999})},\ \Eprint {http://arxiv.org/abs/gr-qc/9909058}
  {arXiv:gr-qc/9909058} \BibitemShut {NoStop}%
\bibitem [{\citenamefont {Li}\ \emph {et~al.}(2021)\citenamefont {Li},
  \citenamefont {Lee}, \citenamefont {Guo},\ and\ \citenamefont
  {Chen}}]{Li:2021zct}%
  \BibitemOpen
  \bibfield  {author} {\bibinfo {author} {\bibfnamefont {Peng-Cheng}\
  \bibnamefont {Li}}, \bibinfo {author} {\bibfnamefont {Tsai-Chen}\
  \bibnamefont {Lee}}, \bibinfo {author} {\bibfnamefont {Minyong}\ \bibnamefont
  {Guo}}, \ and\ \bibinfo {author} {\bibfnamefont {Bin}\ \bibnamefont {Chen}},\
  }\bibfield  {title} {\enquote {\bibinfo {title} {{Correspondence of eikonal
  quasinormal modes and unstable fundamental photon orbits for a Kerr-Newman
  black hole}},}\ }\href {\doibase 10.1103/PhysRevD.104.084044} {\bibfield
  {journal} {\bibinfo  {journal} {Phys. Rev. D}\ }\textbf {\bibinfo {volume}
  {104}},\ \bibinfo {pages} {084044} (\bibinfo {year} {2021})},\ \Eprint
  {http://arxiv.org/abs/2105.14268} {arXiv:2105.14268 [gr-qc]} \BibitemShut
  {NoStop}%
\bibitem [{\citenamefont {Rinc\'on}\ and\ \citenamefont
  {Panotopoulos}(2018)}]{Rincon:2018ktz}%
  \BibitemOpen
  \bibfield  {author} {\bibinfo {author} {\bibfnamefont {\'Angel}\ \bibnamefont
  {Rinc\'on}}\ and\ \bibinfo {author} {\bibfnamefont {Grigoris}\ \bibnamefont
  {Panotopoulos}},\ }\bibfield  {title} {\enquote {\bibinfo {title} {{Greybody
  factors and quasinormal modes for a nonminimally coupled scalar field in a
  cloud of strings in (2+1)-dimensional background}},}\ }\href {\doibase
  10.1140/epjc/s10052-018-6352-5} {\bibfield  {journal} {\bibinfo  {journal}
  {Eur. Phys. J. C}\ }\textbf {\bibinfo {volume} {78}},\ \bibinfo {pages} {858}
  (\bibinfo {year} {2018})},\ \Eprint {http://arxiv.org/abs/1810.08822}
  {arXiv:1810.08822 [gr-qc]} \BibitemShut {NoStop}%
\bibitem [{\citenamefont {Liu}\ \emph {et~al.}(2023)\citenamefont {Liu},
  \citenamefont {Yang}, \citenamefont {\"Ovg\"un}, \citenamefont {Long},\ and\
  \citenamefont {Xu}}]{Liu:2022ygf}%
  \BibitemOpen
  \bibfield  {author} {\bibinfo {author} {\bibfnamefont {Dong}\ \bibnamefont
  {Liu}}, \bibinfo {author} {\bibfnamefont {Yi}~\bibnamefont {Yang}}, \bibinfo
  {author} {\bibfnamefont {Ali}\ \bibnamefont {\"Ovg\"un}}, \bibinfo {author}
  {\bibfnamefont {Zheng-Wen}\ \bibnamefont {Long}}, \ and\ \bibinfo {author}
  {\bibfnamefont {Zhaoyi}\ \bibnamefont {Xu}},\ }\bibfield  {title} {\enquote
  {\bibinfo {title} {{Gravitational ringing and superradiant instabilities of
  the Kerr-like black holes in a dark matter halo}},}\ }\href {\doibase
  10.1140/epjc/s10052-023-11739-w} {\bibfield  {journal} {\bibinfo  {journal}
  {Eur. Phys. J. C}\ }\textbf {\bibinfo {volume} {83}},\ \bibinfo {pages} {565}
  (\bibinfo {year} {2023})},\ \Eprint {http://arxiv.org/abs/2204.11563}
  {arXiv:2204.11563 [gr-qc]} \BibitemShut {NoStop}%
\bibitem [{\citenamefont {Rincon}\ \emph {et~al.}(2022)\citenamefont {Rincon},
  \citenamefont {Gonzalez}, \citenamefont {Panotopoulos}, \citenamefont
  {Saavedra},\ and\ \citenamefont {Vasquez}}]{Rincon:2021gwd}%
  \BibitemOpen
  \bibfield  {author} {\bibinfo {author} {\bibfnamefont {Angel}\ \bibnamefont
  {Rincon}}, \bibinfo {author} {\bibfnamefont {P.~A.}\ \bibnamefont
  {Gonzalez}}, \bibinfo {author} {\bibfnamefont {Grigoris}\ \bibnamefont
  {Panotopoulos}}, \bibinfo {author} {\bibfnamefont {Joel}\ \bibnamefont
  {Saavedra}}, \ and\ \bibinfo {author} {\bibfnamefont {Yerko}\ \bibnamefont
  {Vasquez}},\ }\bibfield  {title} {\enquote {\bibinfo {title} {{Quasinormal
  modes for a non-minimally coupled scalar field in a five-dimensional
  Einstein\textendash{}Power\textendash{}Maxwell background}},}\ }\href
  {\doibase 10.1140/epjp/s13360-022-03438-4} {\bibfield  {journal} {\bibinfo
  {journal} {Eur. Phys. J. Plus}\ }\textbf {\bibinfo {volume} {137}},\ \bibinfo
  {pages} {1278} (\bibinfo {year} {2022})},\ \Eprint
  {http://arxiv.org/abs/2112.04793} {arXiv:2112.04793 [gr-qc]} \BibitemShut
  {NoStop}%
\bibitem [{\citenamefont {Anacleto}\ \emph {et~al.}(2021)\citenamefont
  {Anacleto}, \citenamefont {Campos}, \citenamefont {Brito},\ and\
  \citenamefont {Passos}}]{Anacleto:2021qoe}%
  \BibitemOpen
  \bibfield  {author} {\bibinfo {author} {\bibfnamefont {M.~A.}\ \bibnamefont
  {Anacleto}}, \bibinfo {author} {\bibfnamefont {J.~A.~V.}\ \bibnamefont
  {Campos}}, \bibinfo {author} {\bibfnamefont {F.~A.}\ \bibnamefont {Brito}}, \
  and\ \bibinfo {author} {\bibfnamefont {E.}~\bibnamefont {Passos}},\
  }\bibfield  {title} {\enquote {\bibinfo {title} {{Quasinormal modes and
  shadow of a Schwarzschild black hole with GUP}},}\ }\href {\doibase
  10.1016/j.aop.2021.168662} {\bibfield  {journal} {\bibinfo  {journal} {Annals
  Phys.}\ }\textbf {\bibinfo {volume} {434}},\ \bibinfo {pages} {168662}
  (\bibinfo {year} {2021})},\ \Eprint {http://arxiv.org/abs/2108.04998}
  {arXiv:2108.04998 [gr-qc]} \BibitemShut {NoStop}%
\bibitem [{\citenamefont {Lambiase}\ \emph {et~al.}(2023)\citenamefont
  {Lambiase}, \citenamefont {Pantig}, \citenamefont {Gogoi},\ and\
  \citenamefont {\"Ovg\"un}}]{Lambiase:2023hng}%
  \BibitemOpen
  \bibfield  {author} {\bibinfo {author} {\bibfnamefont {Gaetano}\ \bibnamefont
  {Lambiase}}, \bibinfo {author} {\bibfnamefont {Reggie~C.}\ \bibnamefont
  {Pantig}}, \bibinfo {author} {\bibfnamefont {Dhruba~Jyoti}\ \bibnamefont
  {Gogoi}}, \ and\ \bibinfo {author} {\bibfnamefont {Ali}\ \bibnamefont
  {\"Ovg\"un}},\ }\bibfield  {title} {\enquote {\bibinfo {title}
  {{Investigating the connection between generalized uncertainty principle and
  asymptotically safe gravity in black hole signatures through shadow and
  quasinormal modes}},}\ }\href {\doibase 10.1140/epjc/s10052-023-11848-6}
  {\bibfield  {journal} {\bibinfo  {journal} {Eur. Phys. J. C}\ }\textbf
  {\bibinfo {volume} {83}},\ \bibinfo {pages} {679} (\bibinfo {year} {2023})},\
  \Eprint {http://arxiv.org/abs/2304.00183} {arXiv:2304.00183 [gr-qc]}
  \BibitemShut {NoStop}%
\bibitem [{\citenamefont {Sekhmani}\ and\ \citenamefont
  {Gogoi}(2023)}]{sekhmani_electromagnetic_2023}%
  \BibitemOpen
  \bibfield  {author} {\bibinfo {author} {\bibfnamefont {Yassine}\ \bibnamefont
  {Sekhmani}}\ and\ \bibinfo {author} {\bibfnamefont {Dhruba~Jyoti}\
  \bibnamefont {Gogoi}},\ }\bibfield  {title} {\enquote {\bibinfo {title}
  {Electromagnetic quasinormal modes of dyonic {AdS} black holes with
  quasitopological electromagnetism in a {Horndeski} gravity theory mimicking
  {EGB} gravity at {D} → 4},}\ }\href {\doibase 10.1142/S0219887823501608}
  {\bibfield  {journal} {\bibinfo  {journal} {International Journal of
  Geometric Methods in Modern Physics}\ ,\ \bibinfo {pages} {2350160}}
  (\bibinfo {year} {2023})}\BibitemShut {NoStop}%
\bibitem [{\citenamefont {Gogoi}\ \emph
  {et~al.}(2023{\natexlab{a}})\citenamefont {Gogoi}, \citenamefont
  {\"Ovg\"un},\ and\ \citenamefont {Koussour}}]{Gogoi:2023kjt}%
  \BibitemOpen
  \bibfield  {author} {\bibinfo {author} {\bibfnamefont {Dhruba~Jyoti}\
  \bibnamefont {Gogoi}}, \bibinfo {author} {\bibfnamefont {Ali}\ \bibnamefont
  {\"Ovg\"un}}, \ and\ \bibinfo {author} {\bibfnamefont {M.}~\bibnamefont
  {Koussour}},\ }\bibfield  {title} {\enquote {\bibinfo {title} {{Quasinormal
  modes of black holes in f(Q) gravity}},}\ }\href {\doibase
  10.1140/epjc/s10052-023-11881-5} {\bibfield  {journal} {\bibinfo  {journal}
  {Eur. Phys. J. C}\ }\textbf {\bibinfo {volume} {83}},\ \bibinfo {pages} {700}
  (\bibinfo {year} {2023}{\natexlab{a}})},\ \Eprint
  {http://arxiv.org/abs/2303.07424} {arXiv:2303.07424 [gr-qc]} \BibitemShut
  {NoStop}%
\bibitem [{\citenamefont {Parbin}\ \emph {et~al.}(2023)\citenamefont {Parbin},
  \citenamefont {Gogoi}, \citenamefont {Bora},\ and\ \citenamefont
  {Goswami}}]{Parbin:2022iwt}%
  \BibitemOpen
  \bibfield  {author} {\bibinfo {author} {\bibfnamefont {Nashiba}\ \bibnamefont
  {Parbin}}, \bibinfo {author} {\bibfnamefont {Dhruba~Jyoti}\ \bibnamefont
  {Gogoi}}, \bibinfo {author} {\bibfnamefont {Jyatsnasree}\ \bibnamefont
  {Bora}}, \ and\ \bibinfo {author} {\bibfnamefont {Umananda~Dev}\ \bibnamefont
  {Goswami}},\ }\bibfield  {title} {\enquote {\bibinfo {title} {{Deflection
  angle, quasinormal modes and optical properties of a de Sitter black hole in
  f (T, B) gravity}},}\ }\href {\doibase 10.1016/j.dark.2023.101315} {\bibfield
   {journal} {\bibinfo  {journal} {Phys. Dark Univ.}\ }\textbf {\bibinfo
  {volume} {42}},\ \bibinfo {pages} {101315} (\bibinfo {year} {2023})},\
  \Eprint {http://arxiv.org/abs/2211.02414} {arXiv:2211.02414 [gr-qc]}
  \BibitemShut {NoStop}%
\bibitem [{\citenamefont {Karmakar}\ \emph {et~al.}(2022)\citenamefont
  {Karmakar}, \citenamefont {Gogoi},\ and\ \citenamefont
  {Goswami}}]{karmakar_quasinormal_2022}%
  \BibitemOpen
  \bibfield  {author} {\bibinfo {author} {\bibfnamefont {Ronit}\ \bibnamefont
  {Karmakar}}, \bibinfo {author} {\bibfnamefont {Dhruba~Jyoti}\ \bibnamefont
  {Gogoi}}, \ and\ \bibinfo {author} {\bibfnamefont {Umananda~Dev}\
  \bibnamefont {Goswami}},\ }\bibfield  {title} {\enquote {\bibinfo {title}
  {Quasinormal modes and thermodynamic properties of {GUP}-corrected
  {Schwarzschild} black hole surrounded by quintessence},}\ }\href {\doibase
  10.1142/S0217751X22501809} {\bibfield  {journal} {\bibinfo  {journal}
  {International Journal of Modern Physics A}\ }\textbf {\bibinfo {volume}
  {37}},\ \bibinfo {pages} {2250180} (\bibinfo {year} {2022})}\BibitemShut
  {NoStop}%
\bibitem [{\citenamefont {Gogoi}\ and\ \citenamefont
  {Goswami}(2022)}]{Gogoi:2022wyv}%
  \BibitemOpen
  \bibfield  {author} {\bibinfo {author} {\bibfnamefont {Dhruba~Jyoti}\
  \bibnamefont {Gogoi}}\ and\ \bibinfo {author} {\bibfnamefont {Umananda~Dev}\
  \bibnamefont {Goswami}},\ }\bibfield  {title} {\enquote {\bibinfo {title}
  {{Quasinormal modes and Hawking radiation sparsity of GUP corrected black
  holes in bumblebee gravity with topological defects}},}\ }\href {\doibase
  10.1088/1475-7516/2022/06/029} {\bibfield  {journal} {\bibinfo  {journal}
  {JCAP}\ }\textbf {\bibinfo {volume} {06}},\ \bibinfo {pages} {029} (\bibinfo
  {year} {2022})},\ \Eprint {http://arxiv.org/abs/2203.07594} {arXiv:2203.07594
  [gr-qc]} \BibitemShut {NoStop}%
\bibitem [{\citenamefont {Gogoi}\ \emph
  {et~al.}(2023{\natexlab{b}})\citenamefont {Gogoi}, \citenamefont {Karmakar},\
  and\ \citenamefont {Goswami}}]{Gogoi:2021cbp}%
  \BibitemOpen
  \bibfield  {author} {\bibinfo {author} {\bibfnamefont {Dhruba~Jyoti}\
  \bibnamefont {Gogoi}}, \bibinfo {author} {\bibfnamefont {Ronit}\ \bibnamefont
  {Karmakar}}, \ and\ \bibinfo {author} {\bibfnamefont {Umananda~Dev}\
  \bibnamefont {Goswami}},\ }\bibfield  {title} {\enquote {\bibinfo {title}
  {{Quasinormal modes of nonlinearly charged black holes surrounded by a cloud
  of strings in Rastall gravity}},}\ }\href {\doibase
  10.1142/S021988782350007X} {\bibfield  {journal} {\bibinfo  {journal} {Int.
  J. Geom. Meth. Mod. Phys.}\ }\textbf {\bibinfo {volume} {20}},\ \bibinfo
  {pages} {2350007} (\bibinfo {year} {2023}{\natexlab{b}})},\ \Eprint
  {http://arxiv.org/abs/2111.00854} {arXiv:2111.00854 [gr-qc]} \BibitemShut
  {NoStop}%
\bibitem [{\citenamefont {Gogoi}\ and\ \citenamefont
  {Goswami}(2021)}]{Gogoi:2021dkr}%
  \BibitemOpen
  \bibfield  {author} {\bibinfo {author} {\bibfnamefont {Dhruba~Jyoti}\
  \bibnamefont {Gogoi}}\ and\ \bibinfo {author} {\bibfnamefont {Umananda~Dev}\
  \bibnamefont {Goswami}},\ }\bibfield  {title} {\enquote {\bibinfo {title}
  {{Quasinormal modes of black holes with non-linear-electrodynamic sources in
  Rastall gravity}},}\ }\href {\doibase 10.1016/j.dark.2021.100860} {\bibfield
  {journal} {\bibinfo  {journal} {Phys. Dark Univ.}\ }\textbf {\bibinfo
  {volume} {33}},\ \bibinfo {pages} {100860} (\bibinfo {year} {2021})},\
  \Eprint {http://arxiv.org/abs/2104.13115} {arXiv:2104.13115 [gr-qc]}
  \BibitemShut {NoStop}%
\bibitem [{\citenamefont {Pantig}\ \emph {et~al.}(2022)\citenamefont {Pantig},
  \citenamefont {Mastrototaro}, \citenamefont {Lambiase},\ and\ \citenamefont
  {\"Ovg\"un}}]{Pantig:2022gih}%
  \BibitemOpen
  \bibfield  {author} {\bibinfo {author} {\bibfnamefont {Reggie~C.}\
  \bibnamefont {Pantig}}, \bibinfo {author} {\bibfnamefont {Leonardo}\
  \bibnamefont {Mastrototaro}}, \bibinfo {author} {\bibfnamefont {Gaetano}\
  \bibnamefont {Lambiase}}, \ and\ \bibinfo {author} {\bibfnamefont {Ali}\
  \bibnamefont {\"Ovg\"un}},\ }\bibfield  {title} {\enquote {\bibinfo {title}
  {{Shadow, lensing, quasinormal modes, greybody bounds and neutrino
  propagation by dyonic ModMax black holes}},}\ }\href {\doibase
  10.1140/epjc/s10052-022-11125-y} {\bibfield  {journal} {\bibinfo  {journal}
  {Eur. Phys. J. C}\ }\textbf {\bibinfo {volume} {82}},\ \bibinfo {pages}
  {1155} (\bibinfo {year} {2022})},\ \Eprint {http://arxiv.org/abs/2208.06664}
  {arXiv:2208.06664 [gr-qc]} \BibitemShut {NoStop}%
\bibitem [{\citenamefont {Gogoi}(2024)}]{Gogoi:2024scc}%
  \BibitemOpen
  \bibfield  {author} {\bibinfo {author} {\bibfnamefont {Dhruba~Jyoti}\
  \bibnamefont {Gogoi}},\ }\bibfield  {title} {\enquote {\bibinfo {title}
  {{Violation of Hod\textquoteright{}s conjecture and probing it with optical
  properties of a 5-D black hole in Einstein Gauss\textendash{}Bonnet Bumblebee
  theory of gravity}},}\ }\href {\doibase 10.1016/j.dark.2024.101535}
  {\bibfield  {journal} {\bibinfo  {journal} {Phys. Dark Univ.}\ }\textbf
  {\bibinfo {volume} {45}},\ \bibinfo {pages} {101535} (\bibinfo {year}
  {2024})},\ \Eprint {http://arxiv.org/abs/2405.02455} {arXiv:2405.02455
  [gr-qc]} \BibitemShut {NoStop}%
\bibitem [{\citenamefont {Gogoi}\ and\ \citenamefont
  {Ponglertsakul}(2024)}]{Gogoi:2024vcx}%
  \BibitemOpen
  \bibfield  {author} {\bibinfo {author} {\bibfnamefont {Dhruba~Jyoti}\
  \bibnamefont {Gogoi}}\ and\ \bibinfo {author} {\bibfnamefont {Supakchai}\
  \bibnamefont {Ponglertsakul}},\ }\href@noop {} {\enquote {\bibinfo {title}
  {{Constraints on Quasinormal modes from Black Hole Shadows in regular
  non-minimal Einstein Yang-Mills Gravity}},}\ } (\bibinfo {year} {2024}),\
  \Eprint {http://arxiv.org/abs/2402.06186} {arXiv:2402.06186 [gr-qc]}
  \BibitemShut {NoStop}%
\bibitem [{\citenamefont {Gogoi}\ \emph {et~al.}(2024)\citenamefont {Gogoi},
  \citenamefont {Heidari}, \citenamefont {K̆r\'\i{}̆z},\ and\ \citenamefont
  {Hassanabadi}}]{Gogoi:2023lvw}%
  \BibitemOpen
  \bibfield  {author} {\bibinfo {author} {\bibfnamefont {Dhruba~Jyoti}\
  \bibnamefont {Gogoi}}, \bibinfo {author} {\bibfnamefont {Narges}\
  \bibnamefont {Heidari}}, \bibinfo {author} {\bibfnamefont {Jan}\ \bibnamefont
  {K̆r\'\i{}̆z}}, \ and\ \bibinfo {author} {\bibfnamefont {Hassan}\
  \bibnamefont {Hassanabadi}},\ }\bibfield  {title} {\enquote {\bibinfo {title}
  {{Quasinormal Modes and Greybody Factors of de Sitter Black Holes Surrounded
  by Quintessence in Rastall Gravity}},}\ }\href {\doibase
  10.1002/prop.202300245} {\bibfield  {journal} {\bibinfo  {journal} {Fortsch.
  Phys.}\ }\textbf {\bibinfo {volume} {72}},\ \bibinfo {pages} {2300245}
  (\bibinfo {year} {2024})},\ \Eprint {http://arxiv.org/abs/2307.09976}
  {arXiv:2307.09976 [gr-qc]} \BibitemShut {NoStop}%
\bibitem [{\citenamefont {Gibbons}\ and\ \citenamefont
  {Herdeiro}(2001)}]{Gibbons:2001sx}%
  \BibitemOpen
  \bibfield  {author} {\bibinfo {author} {\bibfnamefont {G.~W.}\ \bibnamefont
  {Gibbons}}\ and\ \bibinfo {author} {\bibfnamefont {C.~A.~R.}\ \bibnamefont
  {Herdeiro}},\ }\bibfield  {title} {\enquote {\bibinfo {title} {{The Melvin
  universe in Born-Infeld theory and other theories of nonlinear
  electrodynamics}},}\ }\href {\doibase 10.1088/0264-9381/18/9/305} {\bibfield
  {journal} {\bibinfo  {journal} {Class. Quant. Grav.}\ }\textbf {\bibinfo
  {volume} {18}},\ \bibinfo {pages} {1677--1690} (\bibinfo {year} {2001})},\
  \Eprint {http://arxiv.org/abs/hep-th/0101229} {arXiv:hep-th/0101229}
  \BibitemShut {NoStop}%
\bibitem [{\citenamefont {Salazar}\ \emph {et~al.}(1987)\citenamefont
  {Salazar}, \citenamefont {Garcia},\ and\ \citenamefont
  {Plebanski}}]{Salazar:1987ap}%
  \BibitemOpen
  \bibfield  {author} {\bibinfo {author} {\bibfnamefont {I.~H.}\ \bibnamefont
  {Salazar}}, \bibinfo {author} {\bibfnamefont {A.}~\bibnamefont {Garcia}}, \
  and\ \bibinfo {author} {\bibfnamefont {J.}~\bibnamefont {Plebanski}},\
  }\bibfield  {title} {\enquote {\bibinfo {title} {{Duality Rotations and Type
  $D$ Solutions to Einstein Equations With Nonlinear Electromagnetic
  Sources}},}\ }\href {\doibase 10.1063/1.527430} {\bibfield  {journal}
  {\bibinfo  {journal} {J. Math. Phys.}\ }\textbf {\bibinfo {volume} {28}},\
  \bibinfo {pages} {2171--2181} (\bibinfo {year} {1987})}\BibitemShut {NoStop}%
\bibitem [{\citenamefont {Gurses}\ and\ \citenamefont
  {Feza}(1975)}]{Gurses:1975vu}%
  \BibitemOpen
  \bibfield  {author} {\bibinfo {author} {\bibfnamefont {Metin}\ \bibnamefont
  {Gurses}}\ and\ \bibinfo {author} {\bibfnamefont {Gursey}\ \bibnamefont
  {Feza}},\ }\bibfield  {title} {\enquote {\bibinfo {title} {{Lorentz Covariant
  Treatment of the Kerr-Schild Metric}},}\ }\href {\doibase 10.1063/1.522480}
  {\bibfield  {journal} {\bibinfo  {journal} {J. Math. Phys.}\ }\textbf
  {\bibinfo {volume} {16}},\ \bibinfo {pages} {2385} (\bibinfo {year}
  {1975})}\BibitemShut {NoStop}%
\bibitem [{\citenamefont {Cui}\ \emph {et~al.}(2023)\citenamefont {Cui} \emph
  {et~al.}}]{Cui:2023uyb}%
  \BibitemOpen
  \bibfield  {author} {\bibinfo {author} {\bibfnamefont {Yuzhu}\ \bibnamefont
  {Cui}} \emph {et~al.},\ }\bibfield  {title} {\enquote {\bibinfo {title}
  {{Precessing jet nozzle connecting to a spinning black hole in M87}},}\
  }\href {\doibase 10.1038/s41586-023-06479-6} {\bibfield  {journal} {\bibinfo
  {journal} {Nature}\ }\textbf {\bibinfo {volume} {621}},\ \bibinfo {pages}
  {711--715} (\bibinfo {year} {2023})},\ \Eprint
  {http://arxiv.org/abs/2310.09015} {arXiv:2310.09015 [astro-ph.HE]}
  \BibitemShut {NoStop}%
\bibitem [{\citenamefont {Slan\'y}\ and\ \citenamefont
  {Stuchl\'\i{}k}(2020)}]{Slany:2020jhs}%
  \BibitemOpen
  \bibfield  {author} {\bibinfo {author} {\bibfnamefont {Petr}\ \bibnamefont
  {Slan\'y}}\ and\ \bibinfo {author} {\bibfnamefont {Zden\v{e}k}\ \bibnamefont
  {Stuchl\'\i{}k}},\ }\bibfield  {title} {\enquote {\bibinfo {title}
  {{Equatorial circular orbits in Kerr\textendash{}Newman\textendash{}de Sitter
  spacetimes}},}\ }\href {\doibase 10.1140/epjc/s10052-020-8142-0} {\bibfield
  {journal} {\bibinfo  {journal} {Eur. Phys. J. C}\ }\textbf {\bibinfo {volume}
  {80}},\ \bibinfo {pages} {587} (\bibinfo {year} {2020})}\BibitemShut
  {NoStop}%
\bibitem [{\citenamefont {Carter}(1968)}]{Carter:1968rr}%
  \BibitemOpen
  \bibfield  {author} {\bibinfo {author} {\bibfnamefont {Brandon}\ \bibnamefont
  {Carter}},\ }\bibfield  {title} {\enquote {\bibinfo {title} {{Global
  structure of the Kerr family of gravitational fields}},}\ }\href {\doibase
  10.1103/PhysRev.174.1559} {\bibfield  {journal} {\bibinfo  {journal} {Phys.
  Rev.}\ }\textbf {\bibinfo {volume} {174}},\ \bibinfo {pages} {1559--1571}
  (\bibinfo {year} {1968})}\BibitemShut {NoStop}%
\bibitem [{\citenamefont {Johannsen}(2013)}]{Johannsen:2013vgc}%
  \BibitemOpen
  \bibfield  {author} {\bibinfo {author} {\bibfnamefont {Tim}\ \bibnamefont
  {Johannsen}},\ }\bibfield  {title} {\enquote {\bibinfo {title} {{Photon Rings
  around Kerr and Kerr-like Black Holes}},}\ }\href {\doibase
  10.1088/0004-637X/777/2/170} {\bibfield  {journal} {\bibinfo  {journal}
  {Astrophys. J.}\ }\textbf {\bibinfo {volume} {777}},\ \bibinfo {pages} {170}
  (\bibinfo {year} {2013})},\ \Eprint {http://arxiv.org/abs/1501.02814}
  {arXiv:1501.02814 [astro-ph.HE]} \BibitemShut {NoStop}%
\bibitem [{\citenamefont {Hioki}\ and\ \citenamefont
  {Maeda}(2009)}]{Hioki:2009na}%
  \BibitemOpen
  \bibfield  {author} {\bibinfo {author} {\bibfnamefont {Kenta}\ \bibnamefont
  {Hioki}}\ and\ \bibinfo {author} {\bibfnamefont {Kei-ichi}\ \bibnamefont
  {Maeda}},\ }\bibfield  {title} {\enquote {\bibinfo {title} {{Measurement of
  the Kerr Spin Parameter by Observation of a Compact Object's Shadow}},}\
  }\href {\doibase 10.1103/PhysRevD.80.024042} {\bibfield  {journal} {\bibinfo
  {journal} {Phys. Rev. D}\ }\textbf {\bibinfo {volume} {80}},\ \bibinfo
  {pages} {024042} (\bibinfo {year} {2009})},\ \Eprint
  {http://arxiv.org/abs/0904.3575} {arXiv:0904.3575 [astro-ph.HE]} \BibitemShut
  {NoStop}%
\bibitem [{\citenamefont {Dymnikova}\ and\ \citenamefont
  {Kraav}(2019)}]{Dymnikova:2019vuz}%
  \BibitemOpen
  \bibfield  {author} {\bibinfo {author} {\bibfnamefont {Irina}\ \bibnamefont
  {Dymnikova}}\ and\ \bibinfo {author} {\bibfnamefont {Kirill}\ \bibnamefont
  {Kraav}},\ }\bibfield  {title} {\enquote {\bibinfo {title} {{Identification
  of a Regular Black Hole by Its Shadow}},}\ }\href {\doibase
  10.3390/universe5070163} {\bibfield  {journal} {\bibinfo  {journal}
  {Universe}\ }\textbf {\bibinfo {volume} {5}},\ \bibinfo {pages} {163}
  (\bibinfo {year} {2019})}\BibitemShut {NoStop}%
\bibitem [{\citenamefont {Iyer}\ and\ \citenamefont
  {Will}(1987)}]{Iyer:1986np}%
  \BibitemOpen
  \bibfield  {author} {\bibinfo {author} {\bibfnamefont {Sai}\ \bibnamefont
  {Iyer}}\ and\ \bibinfo {author} {\bibfnamefont {Clifford~M.}\ \bibnamefont
  {Will}},\ }\bibfield  {title} {\enquote {\bibinfo {title} {{Black Hole Normal
  Modes: A {WKB} Approach. 1. Foundations and Application of a Higher Order
  {WKB} Analysis of Potential Barrier Scattering}},}\ }\href {\doibase
  10.1103/PhysRevD.35.3621} {\bibfield  {journal} {\bibinfo  {journal} {Phys.
  Rev. D}\ }\textbf {\bibinfo {volume} {35}},\ \bibinfo {pages} {3621}
  (\bibinfo {year} {1987})}\BibitemShut {NoStop}%
\bibitem [{\citenamefont {Dias}\ \emph {et~al.}(2022)\citenamefont {Dias},
  \citenamefont {Godazgar},\ and\ \citenamefont {Santos}}]{Dias:2022oqm}%
  \BibitemOpen
  \bibfield  {author} {\bibinfo {author} {\bibfnamefont {Oscar J.~C.}\
  \bibnamefont {Dias}}, \bibinfo {author} {\bibfnamefont {Mahdi}\ \bibnamefont
  {Godazgar}}, \ and\ \bibinfo {author} {\bibfnamefont {Jorge~E.}\ \bibnamefont
  {Santos}},\ }\bibfield  {title} {\enquote {\bibinfo {title} {{Eigenvalue
  repulsions and quasinormal mode spectra of Kerr-Newman: an extended
  study}},}\ }\href {\doibase 10.1007/JHEP07(2022)076} {\bibfield  {journal}
  {\bibinfo  {journal} {JHEP}\ }\textbf {\bibinfo {volume} {07}},\ \bibinfo
  {pages} {076} (\bibinfo {year} {2022})},\ \Eprint
  {http://arxiv.org/abs/2205.13072} {arXiv:2205.13072 [gr-qc]} \BibitemShut
  {NoStop}%
\bibitem [{\citenamefont {Konoplya}\ \emph {et~al.}(2019)\citenamefont
  {Konoplya}, \citenamefont {Zhidenko},\ and\ \citenamefont
  {Zinhailo}}]{Konoplya:2019hlu}%
  \BibitemOpen
  \bibfield  {author} {\bibinfo {author} {\bibfnamefont {R.~A.}\ \bibnamefont
  {Konoplya}}, \bibinfo {author} {\bibfnamefont {A.}~\bibnamefont {Zhidenko}},
  \ and\ \bibinfo {author} {\bibfnamefont {A.~F.}\ \bibnamefont {Zinhailo}},\
  }\bibfield  {title} {\enquote {\bibinfo {title} {{Higher order WKB formula
  for quasinormal modes and grey-body factors: recipes for quick and accurate
  calculations}},}\ }\href {\doibase 10.1088/1361-6382/ab2e25} {\bibfield
  {journal} {\bibinfo  {journal} {Class. Quant. Grav.}\ }\textbf {\bibinfo
  {volume} {36}},\ \bibinfo {pages} {155002} (\bibinfo {year} {2019})},\
  \Eprint {http://arxiv.org/abs/1904.10333} {arXiv:1904.10333 [gr-qc]}
  \BibitemShut {NoStop}%
\bibitem [{\citenamefont {Konoplya}\ and\ \citenamefont
  {Zhidenko}(2011)}]{Konoplya:2011qq}%
  \BibitemOpen
  \bibfield  {author} {\bibinfo {author} {\bibfnamefont {R.~A.}\ \bibnamefont
  {Konoplya}}\ and\ \bibinfo {author} {\bibfnamefont {A.}~\bibnamefont
  {Zhidenko}},\ }\bibfield  {title} {\enquote {\bibinfo {title} {{Quasinormal
  modes of black holes: From astrophysics to string theory}},}\ }\href
  {\doibase 10.1103/RevModPhys.83.793} {\bibfield  {journal} {\bibinfo
  {journal} {Rev. Mod. Phys.}\ }\textbf {\bibinfo {volume} {83}},\ \bibinfo
  {pages} {793--836} (\bibinfo {year} {2011})},\ \Eprint
  {http://arxiv.org/abs/1102.4014} {arXiv:1102.4014 [gr-qc]} \BibitemShut
  {NoStop}%
\bibitem [{\citenamefont {Konoplya}\ and\ \citenamefont
  {Stuchl\'\i{}k}(2017)}]{Konoplya:2017wot}%
  \BibitemOpen
  \bibfield  {author} {\bibinfo {author} {\bibfnamefont {R.~A.}\ \bibnamefont
  {Konoplya}}\ and\ \bibinfo {author} {\bibfnamefont {Z.}~\bibnamefont
  {Stuchl\'\i{}k}},\ }\bibfield  {title} {\enquote {\bibinfo {title} {{Are
  eikonal quasinormal modes linked to the unstable circular null geodesics?}}}\
  }\href {\doibase 10.1016/j.physletb.2017.06.015} {\bibfield  {journal}
  {\bibinfo  {journal} {Phys. Lett. B}\ }\textbf {\bibinfo {volume} {771}},\
  \bibinfo {pages} {597--602} (\bibinfo {year} {2017})},\ \Eprint
  {http://arxiv.org/abs/1705.05928} {arXiv:1705.05928 [gr-qc]} \BibitemShut
  {NoStop}%
\bibitem [{\citenamefont {Konoplya}(2003)}]{Konoplya:2003ii}%
  \BibitemOpen
  \bibfield  {author} {\bibinfo {author} {\bibfnamefont {R.~A.}\ \bibnamefont
  {Konoplya}},\ }\bibfield  {title} {\enquote {\bibinfo {title} {{Quasinormal
  behavior of the d-dimensional Schwarzschild black hole and higher order WKB
  approach}},}\ }\href {\doibase 10.1103/PhysRevD.68.024018} {\bibfield
  {journal} {\bibinfo  {journal} {Phys. Rev. D}\ }\textbf {\bibinfo {volume}
  {68}},\ \bibinfo {pages} {024018} (\bibinfo {year} {2003})},\ \Eprint
  {http://arxiv.org/abs/gr-qc/0303052} {arXiv:gr-qc/0303052} \BibitemShut
  {NoStop}%
\bibitem [{\citenamefont {Yang}\ \emph {et~al.}(2012)\citenamefont {Yang},
  \citenamefont {Nichols}, \citenamefont {Zhang}, \citenamefont {Zimmerman},
  \citenamefont {Zhang},\ and\ \citenamefont {Chen}}]{Yang:2012he}%
  \BibitemOpen
  \bibfield  {author} {\bibinfo {author} {\bibfnamefont {Huan}\ \bibnamefont
  {Yang}}, \bibinfo {author} {\bibfnamefont {David~A.}\ \bibnamefont
  {Nichols}}, \bibinfo {author} {\bibfnamefont {Fan}\ \bibnamefont {Zhang}},
  \bibinfo {author} {\bibfnamefont {Aaron}\ \bibnamefont {Zimmerman}}, \bibinfo
  {author} {\bibfnamefont {Zhongyang}\ \bibnamefont {Zhang}}, \ and\ \bibinfo
  {author} {\bibfnamefont {Yanbei}\ \bibnamefont {Chen}},\ }\bibfield  {title}
  {\enquote {\bibinfo {title} {{Quasinormal-mode spectrum of Kerr black holes
  and its geometric interpretation}},}\ }\href {\doibase
  10.1103/PhysRevD.86.104006} {\bibfield  {journal} {\bibinfo  {journal} {Phys.
  Rev. D}\ }\textbf {\bibinfo {volume} {86}},\ \bibinfo {pages} {104006}
  (\bibinfo {year} {2012})},\ \Eprint {http://arxiv.org/abs/1207.4253}
  {arXiv:1207.4253 [gr-qc]} \BibitemShut {NoStop}%
\bibitem [{\citenamefont {Teukolsky}(1972)}]{Teukolsky:1972my}%
  \BibitemOpen
  \bibfield  {author} {\bibinfo {author} {\bibfnamefont {S.~A.}\ \bibnamefont
  {Teukolsky}},\ }\bibfield  {title} {\enquote {\bibinfo {title} {{Rotating
  black holes - separable wave equations for gravitational and electromagnetic
  perturbations}},}\ }\href {\doibase 10.1103/PhysRevLett.29.1114} {\bibfield
  {journal} {\bibinfo  {journal} {Phys. Rev. Lett.}\ }\textbf {\bibinfo
  {volume} {29}},\ \bibinfo {pages} {1114--1118} (\bibinfo {year}
  {1972})}\BibitemShut {NoStop}%
\bibitem [{\citenamefont {Luna}\ \emph {et~al.}(2023)\citenamefont {Luna},
  \citenamefont {Calder\'on~Bustillo}, \citenamefont {Mart\'\i{}nez},
  \citenamefont {Torres-Forn\'e},\ and\ \citenamefont {Font}}]{Luna:2022rql}%
  \BibitemOpen
  \bibfield  {author} {\bibinfo {author} {\bibfnamefont {Raimon}\ \bibnamefont
  {Luna}}, \bibinfo {author} {\bibfnamefont {Juan}\ \bibnamefont
  {Calder\'on~Bustillo}}, \bibinfo {author} {\bibfnamefont {Juan
  Jos\'e~Seoane}\ \bibnamefont {Mart\'\i{}nez}}, \bibinfo {author}
  {\bibfnamefont {Alejandro}\ \bibnamefont {Torres-Forn\'e}}, \ and\ \bibinfo
  {author} {\bibfnamefont {Jos\'e~A.}\ \bibnamefont {Font}},\ }\bibfield
  {title} {\enquote {\bibinfo {title} {{Solving the Teukolsky equation with
  physics-informed neural networks}},}\ }\href {\doibase
  10.1103/PhysRevD.107.064025} {\bibfield  {journal} {\bibinfo  {journal}
  {Phys. Rev. D}\ }\textbf {\bibinfo {volume} {107}},\ \bibinfo {pages}
  {064025} (\bibinfo {year} {2023})},\ \Eprint
  {http://arxiv.org/abs/2212.06103} {arXiv:2212.06103 [gr-qc]} \BibitemShut
  {NoStop}%
\bibitem [{\citenamefont {Yang}(2021)}]{Yang:2021zqy}%
  \BibitemOpen
  \bibfield  {author} {\bibinfo {author} {\bibfnamefont {Huan}\ \bibnamefont
  {Yang}},\ }\bibfield  {title} {\enquote {\bibinfo {title} {{Relating Black
  Hole Shadow to Quasinormal Modes for Rotating Black Holes}},}\ }\href
  {\doibase 10.1103/PhysRevD.103.084010} {\bibfield  {journal} {\bibinfo
  {journal} {Phys. Rev. D}\ }\textbf {\bibinfo {volume} {103}},\ \bibinfo
  {pages} {084010} (\bibinfo {year} {2021})},\ \Eprint
  {http://arxiv.org/abs/2101.11129} {arXiv:2101.11129 [gr-qc]} \BibitemShut
  {NoStop}%
\bibitem [{\citenamefont {Matyjasek}\ and\ \citenamefont
  {Telecka}(2019)}]{Matyjasek:2019eeu}%
  \BibitemOpen
  \bibfield  {author} {\bibinfo {author} {\bibfnamefont {Jerzy}\ \bibnamefont
  {Matyjasek}}\ and\ \bibinfo {author} {\bibfnamefont {Malgorzata}\
  \bibnamefont {Telecka}},\ }\bibfield  {title} {\enquote {\bibinfo {title}
  {{Quasinormal modes of black holes. II. Pad\'e summation of the higher-order
  WKB terms}},}\ }\href {\doibase 10.1103/PhysRevD.100.124006} {\bibfield
  {journal} {\bibinfo  {journal} {Phys. Rev. D}\ }\textbf {\bibinfo {volume}
  {100}},\ \bibinfo {pages} {124006} (\bibinfo {year} {2019})},\ \Eprint
  {http://arxiv.org/abs/1908.09389} {arXiv:1908.09389 [gr-qc]} \BibitemShut
  {NoStop}%
\bibitem [{\citenamefont {Schutz}\ and\ \citenamefont
  {Will}(1985)}]{Schutz:1985km}%
  \BibitemOpen
  \bibfield  {author} {\bibinfo {author} {\bibfnamefont {Bernard~F.}\
  \bibnamefont {Schutz}}\ and\ \bibinfo {author} {\bibfnamefont {Clifford~M.}\
  \bibnamefont {Will}},\ }\bibfield  {title} {\enquote {\bibinfo {title}
  {{BLACK HOLE NORMAL MODES: A SEMIANALYTIC APPROACH}},}\ }\href {\doibase
  10.1086/184453} {\bibfield  {journal} {\bibinfo  {journal} {Astrophys. J.
  Lett.}\ }\textbf {\bibinfo {volume} {291}},\ \bibinfo {pages} {L33--L36}
  (\bibinfo {year} {1985})}\BibitemShut {NoStop}%
\bibitem [{\citenamefont {Andersson}(1995)}]{Andersson:1995vi}%
  \BibitemOpen
  \bibfield  {author} {\bibinfo {author} {\bibfnamefont {N.}~\bibnamefont
  {Andersson}},\ }\bibfield  {title} {\enquote {\bibinfo {title} {{Scattering
  of massless scalar waves by a Schwarzschild black hole: A Phase integral
  study}},}\ }\href {\doibase 10.1103/PhysRevD.52.1808} {\bibfield  {journal}
  {\bibinfo  {journal} {Phys. Rev. D}\ }\textbf {\bibinfo {volume} {52}},\
  \bibinfo {pages} {1808--1820} (\bibinfo {year} {1995})}\BibitemShut {NoStop}%
\bibitem [{\citenamefont {Cardoso}\ \emph {et~al.}(2009)\citenamefont
  {Cardoso}, \citenamefont {Miranda}, \citenamefont {Berti}, \citenamefont
  {Witek},\ and\ \citenamefont {Zanchin}}]{Cardoso:2008bp}%
  \BibitemOpen
  \bibfield  {author} {\bibinfo {author} {\bibfnamefont {Vitor}\ \bibnamefont
  {Cardoso}}, \bibinfo {author} {\bibfnamefont {Alex~S.}\ \bibnamefont
  {Miranda}}, \bibinfo {author} {\bibfnamefont {Emanuele}\ \bibnamefont
  {Berti}}, \bibinfo {author} {\bibfnamefont {Helvi}\ \bibnamefont {Witek}}, \
  and\ \bibinfo {author} {\bibfnamefont {Vilson~T.}\ \bibnamefont {Zanchin}},\
  }\bibfield  {title} {\enquote {\bibinfo {title} {{Geodesic stability,
  Lyapunov exponents and quasinormal modes}},}\ }\href {\doibase
  10.1103/PhysRevD.79.064016} {\bibfield  {journal} {\bibinfo  {journal} {Phys.
  Rev. D}\ }\textbf {\bibinfo {volume} {79}},\ \bibinfo {pages} {064016}
  (\bibinfo {year} {2009})},\ \Eprint {http://arxiv.org/abs/0812.1806}
  {arXiv:0812.1806 [hep-th]} \BibitemShut {NoStop}%
\end{thebibliography}%

\end{document}